\documentclass[11pt]{article}
\usepackage[dvipdfmx]{graphicx}
\usepackage{amsmath}
\usepackage{amssymb}
\usepackage{latexsym}
\usepackage{bm}
\usepackage{overpic}
\usepackage[dvips]{color}
\usepackage{cases}
\usepackage{lscape}
\usepackage{ulem}
\usepackage{float}
\usepackage{fancybox}

\newcommand{\ds}{\displaystyle}
\newtheorem{theorem}{Theorem}
\newtheorem{lemma}{Lemma}
\newtheorem{remark}{Remark}
\newtheorem{example}{Example}

\begin{document}
%

\baselineskip 6mm

\begin{center}
{\LARGE\bf A Sequential Computation Algorithm for the Center\\[2mm]
of the Smallest Enclosing Ball}\\[5mm]
\large Kenji~Nakagawa\,$^\ast$ and Yoshinori~Takei\,${^\dagger}$\\[5mm]
\end{center}
\footnote[0]{$^\ast$Nagaoka University of Technology, e-mail:nakagawa@vos.nagaokaut.ac.jp,\\
\hspace{6mm}${^\dagger}$independent researcher}
\begin{abstract}
In this paper, we consider the problem of finding the center $Q^\ast$ of the SEB (smallest enclosing ball) for $n$ points in $d$-dimensional Euclidean space. One application of the SEB is SVDD (support vector data description) in support vector machines. Our objective is to develop a sequential computation algorithm for determining the barycentric coordinate of $Q^\ast$. To achieve it, we apply the concept of the Arimoto-Blahut algorithm, which is a sequential computation algorithm used to compute the channel capacity. We first consider the case in which an equidistant point $\widetilde{Q}$ from the $n$ points exists, and construct a recurrence formula that converges to the barycentric coordinate $\widetilde{\bm\lambda}$ of $\widetilde{Q}$. When $\widetilde{Q}$ lies within the convex hull of the $n$ points, $\widetilde{Q}$ coincides with $Q^\ast$, hence in this case, the recurrence formula converges to the barycentric coordinate $\bm\lambda^\ast$ of $Q^\ast$. The resulting recurrence formula is very simple because it uses only the coordinates of the $n$ points. The computational complexity, with an approximation error of $\epsilon$ to the exact solution $\widetilde{\bm\lambda}$, is $O(\kappa n^2\log(1/\epsilon))$, where $\kappa$ is a value determined by the $n$ points. Furthermore, we modify the algorithm so that it can also be applied in cases where $\widetilde{Q}$ does not exist, and evaluate the convergence performance numerically. We compare the proposed algorithm with conventional algorithms in terms of run time and computational accuracy through several examples. The proposed algorithm has some advantages and some disadvantages compared to the conventional algorithms, but overall, since the proposed algorithm can be computed using a very simple formula, it is considered sufficiently practical.
\end{abstract}


\section{Introduction}
The SEB (smallest enclosing ball) for $n$ points $P^1, \dots, P^n$ in $d$-dimensional Euclidean space $\mathbb{R}^d$ is the smallest-radius ball that contains all points $P^1,\dots,P^n$ either inside or on its surface. The center $Q^\ast$ of the SEB is determined by the $\min$-$\max$ optimization problem
\begin{align}
\Gamma\triangleq\min_{Q\in{\mathbb R}^d}\max_{1\leq i\leq n}\|P^i-Q\|^2,\label{eqn:Gammadefinition}
\end{align}
i.e., $Q=Q^\ast$ is the point that minimizes the maximum squared distance from all given points and $\sqrt{\Gamma}$ is the radius of the SEB. Our objective is to develop a sequential computation algorithm to determine $Q^\ast$ using the coordinates of the given points $P^1,\dots,P^n$.

One application of the SEB is SVDD (support vector data description) in support vector machines \cite{tax}. In SVDD, the training data consist of points $P^1,\dots,P^n$, and the SEB is used to characterize these training data points.
\subsection{Sequential Computation Algorithm in this Study}
The purpose of this study is to develop a sequence $\{Q^N\}_{N=0,1,\dots}$ that converges to the center $Q^\ast$ of the SEB. Specifically, we introduce a recurrence formula for the barycentric coordinates $\bm{\lambda}^N$ of $Q^N$ with respect to the given points $P^1,\dots,P^n$, and use $\bm{\lambda}^N$ to construct $Q^N$. To achieve it, we referred to idea of the Arimoto-Blahut algorithm \cite{ari},\cite{bla}. By adapting its basic concept, we establish an efficient sequential algorithm for $\bm\lambda^N$.
\subsubsection{Arimoto-Blahut Algorithm}
The Arimoto-Blahut algorithm is a sequential computation algorithm used to compute the channel capacity $C$ of a discrete memoryless channel \cite{ari},\cite{bla}. In the case of a channel with $n$ input symbols and $d$ output symbols, let $\Delta^n$ be the set of input probability distributions $\bm{\lambda}$ and $I(\bm\lambda)$ be the mutual information for $\bm\lambda$, then the channel capacity $C$ is defined \cite{csi},\cite{gal} by
\begin{align}
C\triangleq\max_{\bm\lambda \in \Delta^n}I(\bm\lambda).\label{eqn:Cdefinition}
\end{align}

The channel capacity $C$ is also formulated as a $\min$-$\max$ problem as follows. Let $P^i\in\Delta^d,\,i=1,\dots,n$ be the output distribution when the $i$-th input symbol is transmitted, where $\Delta^d$ is the set of output distributions. Further, let $D(P^i\|Q),\,Q\in\Delta^d$ be the Kullback-Leibler divergence \cite{csi},\cite{gal} of $P^i$ and $Q$, then it holds \cite{csi} that
\begin{align}
C=\min_{Q\in\Delta^d}\max_{1\leq i\leq n}D(P^i\|Q).\label{eqn:minmaxofKLD}
\end{align}

In (\ref{eqn:minmaxofKLD}), by replacing the probability space $\Delta^d$ with Euclidean space $\mathbb{R}^d$ and replacing the Kullback-Leibler divergence $D(P^i\|Q)$ with the squared Euclidean distance $\|P^i-Q\|^2$, we obtain the problem (\ref{eqn:Gammadefinition}) of finding the SEB.

Now, for $\bm\lambda\triangleq{^t}(\lambda_1,\dots,\lambda_n)\in\Delta^n$ and $P^i\triangleq{^t}(P^i_1,\dots,P^i_d)\in\mathbb{R}^d,\,i=1,\dots,n$, function $J(\bm\lambda)$, which is analogous to the mutual information $I(\bm\lambda)$, is defined \cite{sch} as follows.
\begin{align}
J(\bm\lambda)=\sum_{i=1}^n\left(\sum_{j=1}^d(P^i_j)^2\right)\lambda_i-\sum_{j=1}^d\left(\sum_{i=1}^n P^i_j\lambda_i\right)^2,\,\bm{\lambda}\in\Delta^n.\label{eqn:Jdefinition}
\end{align}
$J(\bm\lambda)$ is a non-negative and concave function like the mutual information. Then, the optimization problem (\ref{eqn:Gammadefinition}) for the SEB becomes \cite{sch}
\begin{align}
\Gamma=\max_{\bm\lambda\in\Delta^n}J(\bm\lambda).\label{eqn:Gamma=maxJ}
\end{align}

In this study, we develop a sequential computation algorithm to determine the barycentric coordinate of the center of the SEB based on the above similarity.

In the Arimoto-Blahut algorithm \cite{ari},\cite{bla}, the maximization of the mutual information $I(\bm\lambda)$ in (\ref{eqn:Cdefinition}) is changed to a two-variable problem involving $\bm{\lambda}\in\Delta^n$ and a new variable $\varphi$ in a set $\Delta$ of probability distributions, resulting in
\begin{align}
C=\max_{\bm\lambda\in\Delta^n}\max_{\varphi\in\Delta}I(\bm\lambda,\varphi).\label{eqn:Ilambdavarphi}
\end{align}
This two-variable optimization is solved alternately for $\bm{\lambda}$ and $\varphi$, yielding a sequential computation algorithm. This approach is a special case of alternating minimization procedures in \cite{csi2}. There is also a case that the EM (expectation-maximization) algorithm used in statistics is explained using this framework.

\subsection{Achievements of this study}
We begin by assuming a condition that the points $P^1,\dots,P^n$ are in general position. This condition is referred to as the rank condition (11) in Section \ref{sub:rankcondition}. Under this assumption, we prove in Theorem \ref{the:barycentriccoordinate} that there exist a unique point $\widetilde{Q}$ that is equidistant from $P^1,\dots,P^n$, and a unique barycentric coordinate $\widetilde{\bm\lambda}$ of $\widetilde{Q}$ with respect to $P^1,\dots,P^n$. In Section \ref{sec:RecurrenceFormulaforlambdaN}, we construct a recurrence formula $\bm\lambda^N$ for the barycentric coordinates, and in Theorem \ref{the:mainsecond}, we show that $\bm\lambda^N$ converges to $\widetilde{\bm\lambda}$ as $N\to\infty$ and evaluate the convergence rate. In Theorem \ref{the:computationalcomplexity}, we evaluate the computational complexity of the recurrence formula $\bm\lambda^N$. Let $\bm\lambda^\ast$ denote the barycentric coordinate of the center $Q^\ast$ of the SEB. If $\widetilde{\bm\lambda}\in\Delta^n$, then $\widetilde{\bm\lambda}=\bm\lambda^\ast$ holds \cite[Lemma 2]{FGK03}, and thus $\bm\lambda^N$ is a recurrence formula that converges to $\bm\lambda^\ast$.

The characteristics of the obtained algorithm are as follows.

- Simple Computation: The obtained recurrence formula is based on simple operations using only the coordinates of the given $n$ points, hence the algorithm is easier to implement than conventional methods.

- Computational Complexity: When the approximation error is $\epsilon$, the computational complexity of the proposed algorithm is $O(\kappa n^2\log(1/\epsilon))$ under the rank condition (\ref{eqn:rankdefinition}), where $\kappa$ is a value determined by the $n$ points.

Next, we consider the case where the rank condition (\ref{eqn:rankdefinition}) holds but $\widetilde{\bm\lambda}\notin\Delta^n$, as well as the case where the rank condition (\ref{eqn:rankdefinition}) does not hold. For these cases, in Section \ref{sec:heuristic}, we propose a heuristic algorithm by modifying the recurrence formula $\bm\lambda^N$ obtained above, and numerically evaluate its convergence performance. 

We compare the proposed algorithm with conventional algorithms in terms of run time and computational accuracy through several examples. The proposed algorithm has some advantages and some disadvantages compared to the conventional algorithms, but overall, since the proposed algorithm can be computed using a very simple formula, it is considered sufficiently practical.

\section{Previous Research}
Numerous algorithms have been proposed for computing the SEB. One well-known exact algorithm, meaning an algorithm that gives the exact solution, for the SEB is by Welzl \cite{Welzl}. Given finite point sets $\mathcal{P}, \mathcal{Q} \subset \mathbb{R}^d$, $\mathrm{mb}(\mathcal{P}, \mathcal{Q})$ denotes the smallest $d$-dimensional ball that encloses $\mathcal{P}$ while containing $\mathcal{Q}$ on its surface. The points in $\mathcal{Q}$ are said to {\it support} the enclosing ball. For any $p\in\mathcal{P}$, if $p\in\mathrm{mb}(\mathcal{P}\setminus\{p\},\mathcal{Q})$, then $\mathrm{mb}(\mathcal{P},\mathcal{Q})=\mathrm{mb}(\mathcal{P}\setminus\{p\},\mathcal{Q})$. Otherwise, if $p\notin\mathrm{mb}(\mathcal{P}\setminus \{p\},\mathcal{Q})$, then $\mathrm{mb}(\mathcal{P},\mathcal{Q})=\mathrm{mb}(\mathcal{P}\setminus\{p\},\mathcal{Q}\cup\{p\})$. Welzl \cite{Welzl} begins with $\mathcal{P}=\{P^1,\dots,P^n\},\mathcal{Q}=\emptyset$ and updates $(\mathcal{P},\mathcal{Q})$ recursively until either $\mathcal{P}=\emptyset$ or $|\mathcal{Q}|=d+1$, then $\mathrm{mb}(\mathcal{P},\mathcal{Q})$ can be computed directly. Since the set $\mathcal{Q}$ of supporting points is explicitly maintained and explored, an exact solution is obtained. The selection of $p\in\mathcal{P}$ affects the run time, but by randomly permuting $P^1,\dots,P^n$ initially, the expected run time depends linearly on $n$. However, the dependency on the dimension $d$ is significant, where an upper bound of the expected run time is $O(n(d+1)(d+1)!)$. This method will be referred to as Welzl's method in Section \ref{sec:examples1}.  

Another exact algorithm for searching the support set $\mathcal{Q}$ of the SEB was proposed in FGK \cite{FGK03}. This method performs ``dropping'' and ``walking'', repeatedly. Dropping is a process of removing points from $\mathcal{Q}$. If the barycentric coordinate of the center of the tentative enclosing ball has negative values, then the corresponding points of $\mathcal{Q}$ to those negative values are removed from $\mathcal{Q}$. Walking is a process of adding a point to $\mathcal{Q}$. If reducing the tentative enclosing ball after dropping first encounters a point, then the point is added to $\mathcal{Q}$. While FGK \cite{FGK03} does not provide an upper bound on the theoretical run time, experimental results show that it solves a uniformly random instance with $d = 2000, n = 2000$ in about 200 seconds on a 480MHz SUN4 computer. This method will be referred to as FGK's method in Section \ref{sec:examples1}.

Several approximation algorithms have been proposed that output a ball enclosing the given finite point set $\mathcal{P}$, with a radius at most $1+\varepsilon$ times that of the SEB $\mathrm{mb}(\mathcal{P}, \emptyset)$. Among these are core-set-based methods \cite{BHI},\cite{KMY}. An $\varepsilon$-core-set $\mathcal{S}$ of a point set $\mathcal{P} \subset \mathbb{R}^d$ is a subset $\mathcal{S} \subset \mathcal{P}$ such that the radius of its SEB is at least $1/(1+\varepsilon)$ times the radius of the SEB of $\mathcal{P}$ \cite{BHI}. In \cite{KMY}, an algorithm based on a core-set of size $O(1/\varepsilon)$ achieves a run time of $O(nd/\varepsilon + (d^2/\varepsilon^{3/2})(\varepsilon^{-1}+d) \log(1/\varepsilon))$. Also, core-set-based algorithms that achieve run time of $O(nd/\varepsilon)$ have been proposed \cite{Clarkson},\cite{Panigrahy},\cite{Yildirim}.

Meanwhile, \cite{ALY} solves a saddle-point optimization problem of the form $\max_{x\in\mathbb{R}^d}\min_{y\in\Delta^n}$
$(1/n)\,{^t}yAx+(1/n)\,{^t}yb+\lambda H(y)-(\gamma/2)\|x\|_2^2$, where $\lambda,\gamma>0$ are stabilization parameters and $H(y)=\sum_{i=1}^{n}y_i\log y_i$. They \cite{ALY} used a stochastic sequential update method for $x,y$ achieving a run time of $\widetilde{O}(nd+n\sqrt{d/\varepsilon})$, where $\widetilde{O}$ denotes $O$ notation ignoring polynomial factors of $\log$.  

A method based on randomized online gradient descent, the multiplicative weight technique \cite{CHW} runs in $\widetilde{O}(n/\varepsilon^2+d/\varepsilon)$ time, which is sublinear in $nd$, and outputs a $(1+\varepsilon)$-approximate solution with probability at least $1/2$. Ding \cite{DING} further improves the run time under a stability condition, which means that removing a small number of points does not significantly reduce the radius of the SEB. Additionally, \cite{DING} introduces many related studies that could not be covered here.  

In our previous work \cite{nak2}, we developed a method to compute the center $Q^*$ of the SEB under certain restrictive conditions by performing a finite sequence of projections from equidistant point onto affine subspaces.
\section{Formulation of Smallest Enclosing Ball Problem}

In this section, we formulate SEB problem in Euclidean space $\mathbb{R}^d$ and define notations used in this paper. Throughout this paper, we primarily consider vectors as column vectors. Hence, for example, we should write as
\begin{align}
P=\begin{pmatrix}
P_1\\\vdots\\P_d
\end{pmatrix}\in{\mathbb R}^d,
\end{align}
however, to save space, we represent it as a transposed row vector: $P={^t}(P_1,\dots,P_d)\in\mathbb{R}^d$.  

Now, we consider $n$ points $P^1,\dots,P^n$ in $\mathbb{R}^d$, where their coordinates are given by $P^i={^t}(P^i_1,\dots,P^i_d)\in\mathbb{R}^d,\,i=1, \dots, n$. We define a matrix $\Phi$ as
\begin{align}
\Phi\triangleq\begin{pmatrix}
P^1,\dots,P^n
\end{pmatrix}
=
\begin{pmatrix}
P^1_1 & \dots  & P^n_1\\
\vdots & & \vdots\\
P^1_d & \dots & P^n_d
\end{pmatrix}\in{\mathbb R}^{d\times n}.
\end{align}

For points $P={^t}(P_1,\dots,P_d)$ and $Q={^t}(Q_1,\dots,Q_d)\in\mathbb{R}^d$, we define the inner product as $(P,Q)=\sum_{j=1}^dP_jQ_j$, the Euclidean norm as $\|P\|=\{\sum_{j=1}^{d} (P_j)^2\}^{1/2}$, and the Euclidean distance as $\|P-Q\|=\{\sum_{j=1}^{d}(P_j-Q_j)^2\}^{1/2}$.
\subsection{Definition of Smallest Enclosing Ball}
{\it The SEB (smallest enclosing ball)} for the $n$ points $P^1,\dots,P^n$ in $\mathbb{R}^d$ is the smallest-radius $d$-dimensional ball that encloses all of $P^1,\dots,P^n$ either inside or on its surface. That is, $Q=Q^\ast$ that achieves
\begin{align}
\Gamma\triangleq\min_{Q\in{\mathbb R}^d}\max_{1\leq i\leq n}\|P^i-Q\|^2\label{eqn:Gammadefinition2}
\end{align}
is the center of the SEB, and $\sqrt{\Gamma}$ is its radius.
\subsection{Definition of Equidistant Point}
If there exists a point equidistant from all $n$ points $P^1,\dots,P^n$, we denote this point as $\widetilde{Q}$, i.e., $\widetilde{Q}$ satisfies
\begin{align}
\|P^i-\widetilde{Q}\|=\|P^{i'}-\widetilde{Q}\|,\ 1\leq i,i'\leq n.
\end{align}
We call $\widetilde{Q}$ the {\it equidistant point} from $P^1,\dots,P^n$, or the circumcenter of the point set $\{P^1,\dots,P^n\}$.
\subsection{Example: Relation Between Center $Q^\ast$ of SEB and Equidistant Point $\widetilde{Q}$}

\begin{figure}[H]
\centering
\begin{overpic}[width=0.45\columnwidth]{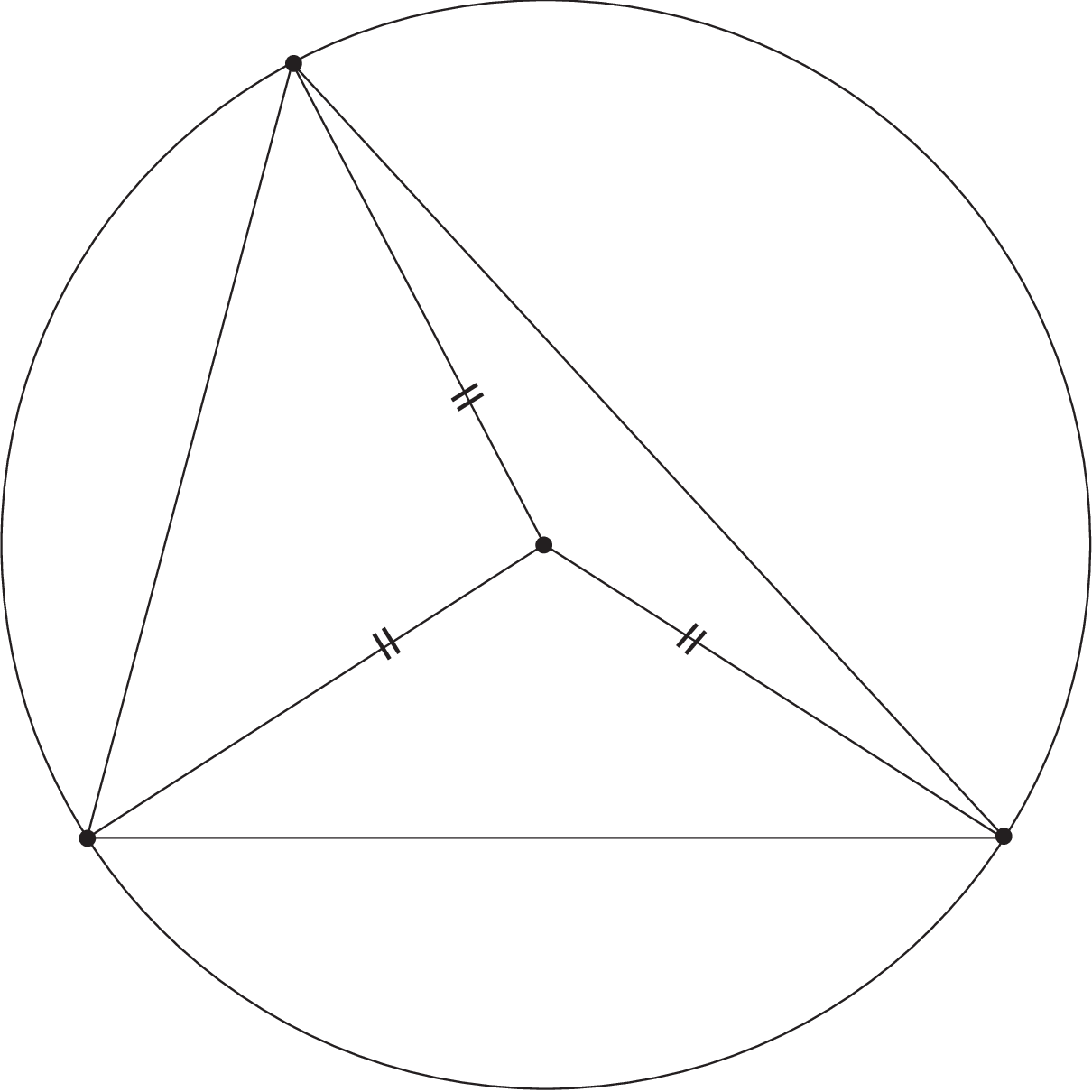}
\put(60,70){$Q^\ast=\widetilde{Q}$}
\put(59,67){\vector(-1,-2){8}}
\put(0,19){$P^1$}
\put(93,19){$P^2$}
\put(24,96){$P^3$}
\end{overpic}
\caption{$\triangle P^1P^2P^3$ is an acute triangle.}
\label{fig:circle1}
\end{figure}

\begin{figure}[H]
\centering
\begin{overpic}[width=0.45\columnwidth]{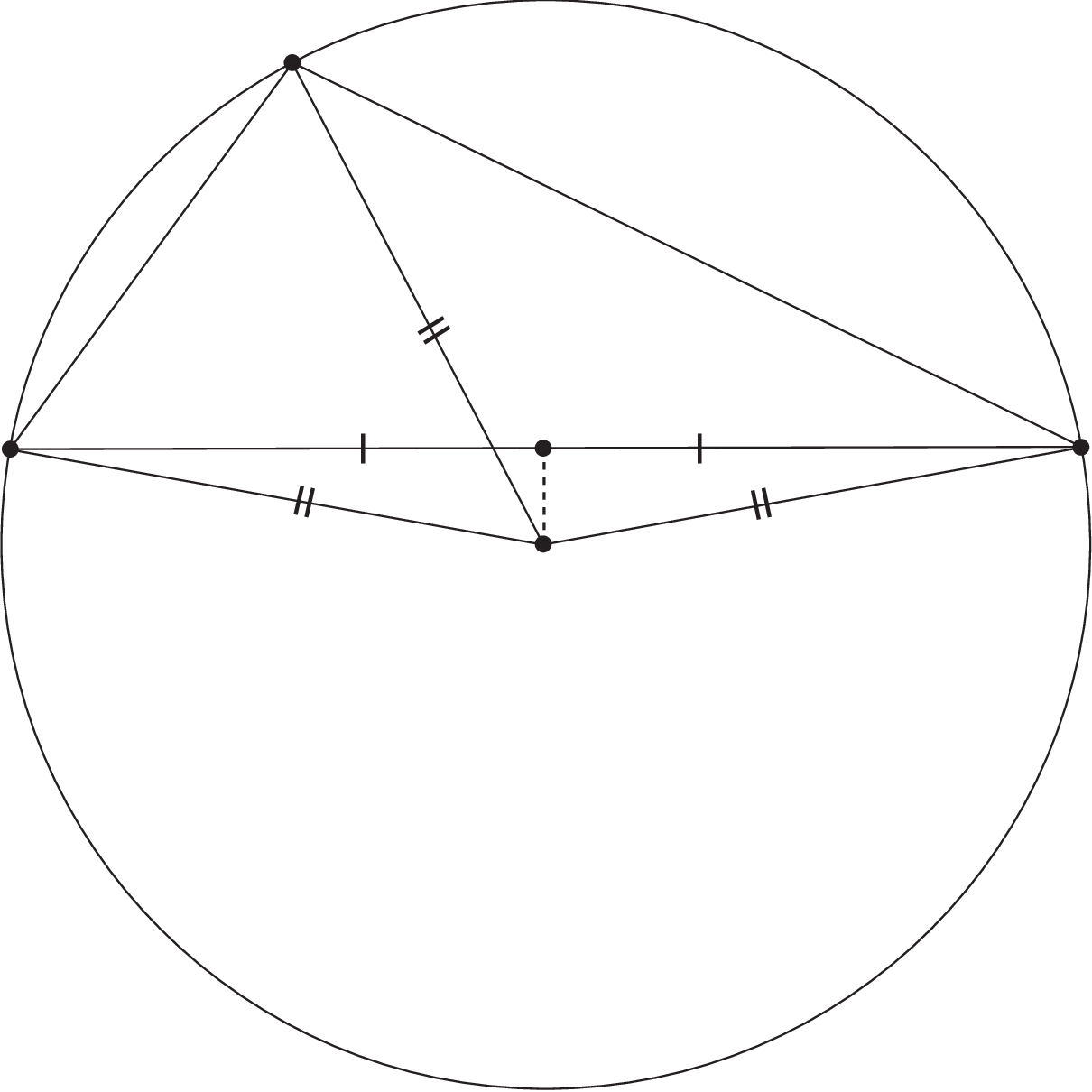}
\put(48,42){$\widetilde{Q}$}
\put(48,63){$Q^\ast$}
\put(-7,56){$P^1$}
\put(100,56){$P^2$}
\put(23,96){$P^3$}
\end{overpic}
\caption{$\triangle P^1P^2P^3$ is an obtuse triangle.}
\label{fig:circle2}
\end{figure}

\begin{figure}[H]
\centering
\begin{overpic}[width=0.45\columnwidth]{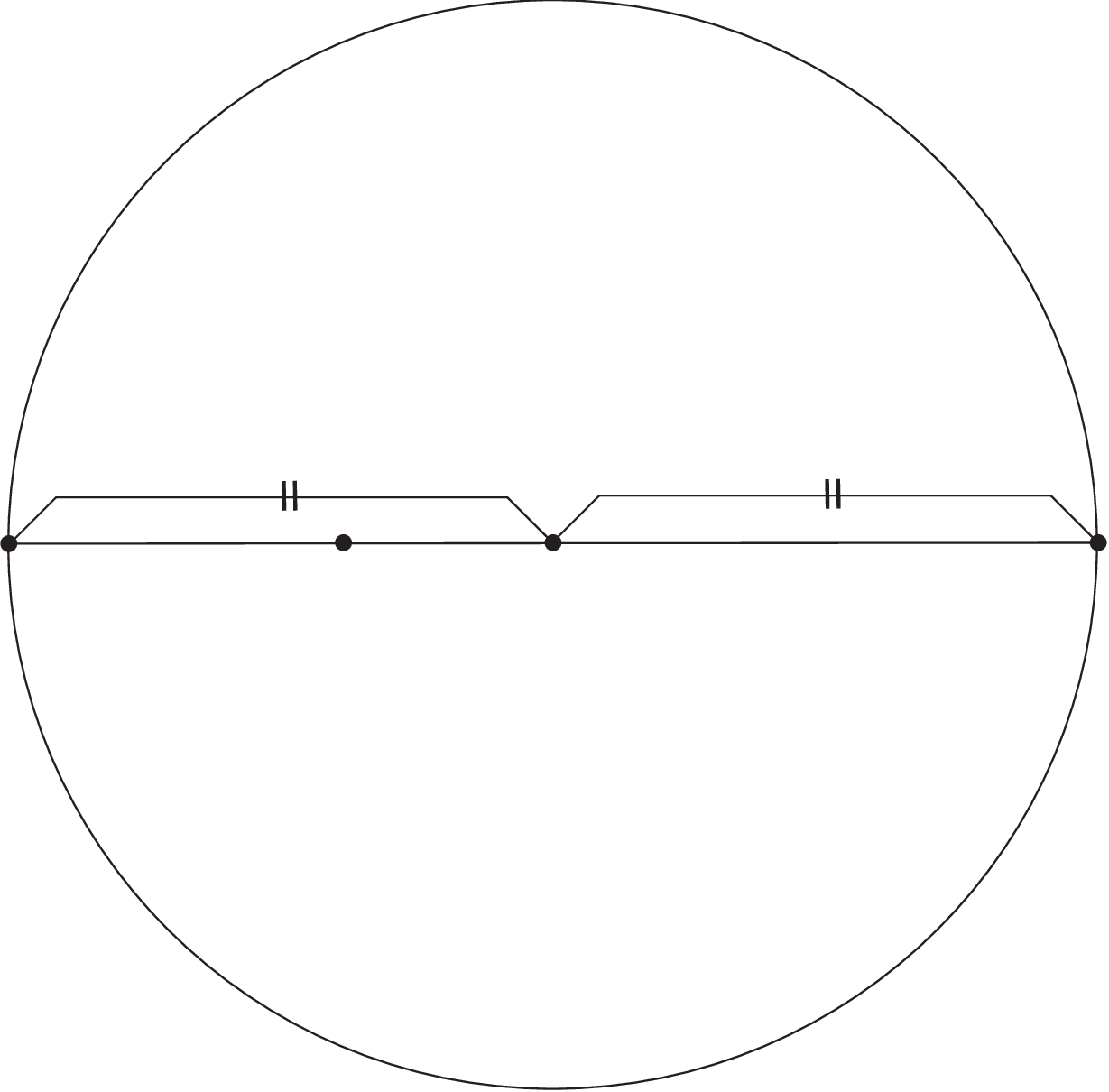}
\put(47,43){$Q^\ast$}
\put(-7,42){$P^1$}
\put(100,43){$P^2$}
\put(29,42){$P^3$}
\end{overpic}
\caption{Three points $P^1,\,P^2,\,P^3$ are collinear.}
\label{fig:circle3}
\end{figure}
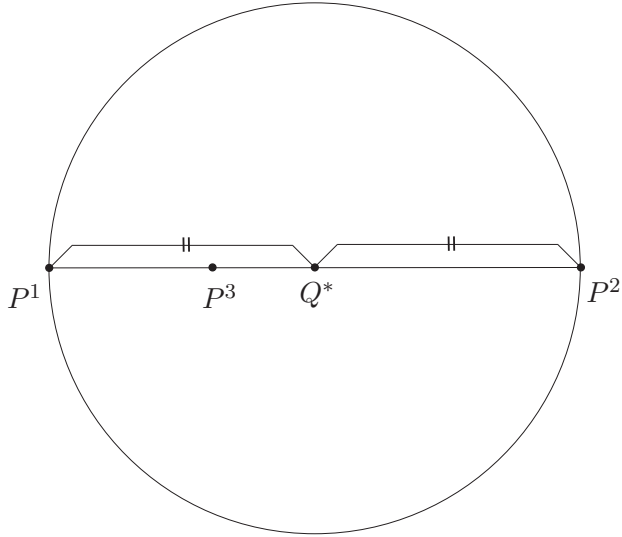

Consider the smallest enclosing circle (ball) for three points $P^1, P^2, P^3$ in $\mathbb{R}^2$. There are three possible cases:
\begin{itemize}
\item[(i)] $\triangle P^1P^2P^3$ is an acute triangle. See Figure \ref{fig:circle1}.

In this case, the equidistant point $\widetilde{Q}$ from $P^1, P^2, P^3$ exists and coincides with the center of the smallest enclosing circle, i.e., $Q^\ast = \widetilde{Q}$.
\item[(ii)] $\triangle P^1P^2P^3$ is an obtuse triangle. See Figure \ref{fig:circle2}.

In this case, the equidistant point $\widetilde{Q}$ from $P^1, P^2, P^3$ exists, but $Q^\ast \neq \widetilde{Q}$. The center $Q^\ast$ is the equidistant point from $P^1$ and $P^2$, excluding the obtuse-angled vertex $P^3$. According to Theorem \ref{the:minlambdaPhitildeQ} below, $Q^\ast$ is the point that achieves $\min_{Q\in\triangle P^1P^2P^3} \|Q-\widetilde{Q}\|^2$.
\item[(iii)] The three points $P^1, P^2, P^3$ are collinear. See Figure \ref{fig:circle3}.

In this case, there is no equidistant point $\widetilde{Q}$ from $P^1, P^2, P^3$. The center $Q^\ast$ is the equidistant point from $P^1$ and $P^2$.
\end{itemize}

\bigskip
\bigskip


In the above cases, in (i) and (ii), we have rank\,$(P^2 - P^1, P^3 - P^1)=2$, and in (iii), rank\,$(P^2-P^1, P^3-P^1)=1$.
\subsection{Rank Condition}
\label{sub:rankcondition}
Consider the case where the points $P^1, \dots, P^n$ are in general position, meaning that
\begin{align}
\text{rank}\,\begin{pmatrix}P^2-P^1,\dots,P^n-P^1\end{pmatrix}=n-1\label{eqn:rankdefinition}
\end{align}
holds. We refer to this condition as the {\it rank condition}.

Under the rank condition (\ref{eqn:rankdefinition}), we will prove that there exists an equidistant point $\widetilde{Q}$ from $P^1, \dots, P^n$, and that the barycentric coordinate of $\widetilde{Q}$ exists. We then construct a recurrence formula for the barycentric coordinates, analyze its convergence, and evaluate its convergence rate. After that, we will also examine cases where the rank condition (\ref{eqn:rankdefinition}) does not hold.
\subsection{Barycentric Coordinate}
First, we define a set $L^n \subset \mathbb{R}^n$ as
\begin{align}
L^n\triangleq\left\{\bm\lambda={^t}(\lambda_1,\dots,\lambda_n)\in{\mathbb R}^n\,\middle|\,\sum_{i=1}^n\lambda_i=1\right\}\subset{\mathbb R}^n.
\end{align}
If a point $Q\in\mathbb{R}^d$ can be uniquely expressed as $Q=\Phi \bm{\lambda}$ for some $\bm{\lambda} \in L^n$, then $\bm{\lambda}$ is called the {\it barycentric coordinate} of $Q$ with respect to $P^1,\dots,P^n$.

Next, we define a set $\Phi L^n\subset\mathbb{R}^d$ as
\begin{align}
\Phi L^n\triangleq\left\{\Phi\bm\lambda\,\middle|\,\bm\lambda\in L^n\right\}\subset{\mathbb R}^d.
\end{align}
We have the following lemma.
\begin{lemma}
\label{lem:barycentriccoordinateisunique}
Assume the rank condition (\ref{eqn:rankdefinition}). For $Q\in\Phi L^n$, the representation $Q=\Phi\bm{\lambda}$ is unique for $\bm{\lambda}={^t}(\lambda_1,\dots,\lambda_n)\in L^n$. That is, $\bm{\lambda}$ is the barycentric coordinate of $Q$ with respect to $P^1,\dots,P^n$.
\end{lemma}
\noindent{\bf Proof:} If there exists another $\bm{\lambda}' = {^t}(\lambda'_1, \dots, \lambda'_n) \in L^n$ such that $Q=\sum_{i=1}^nP^i\lambda_i=\sum_{i=1}^nP^i\lambda'_i$, then, since $\sum_{i=1}^n \lambda_i=\sum_{i=1}^n\lambda'_i=1$, it follows that $\sum_{i=2}^n(P^i-P^1)\lambda_i=\sum_{i=2}^n(P^i-P^1)\lambda'_i$. By the rank condition (\ref{eqn:rankdefinition}), the vectors $P^i - P^1,\,i=2,\dots,n$ are linearly independent, implying that $\bm{\lambda} = \bm{\lambda}'$. $\hfill\square$

\bigskip

Now, we define a set $\Delta^n\subset L^n$ as
\begin{align}
\Delta^n \triangleq \left\{ \bm{\lambda} = {^t}(\lambda_1, \dots, \lambda_n) \in \mathbb{R}^n\,\middle|\,\sum_{i=1}^n \lambda_i = 1, \ \lambda_i \geq 0,\ i=1,\dots,n \right\} \subset L^n.
\end{align}
Then, we define a set $\Phi\Delta^n\subset\mathbb R^d$ as
\begin{align}
\Phi\Delta^n \triangleq \left\{ \Phi \bm{\lambda}\,\middle|\,\bm{\lambda} \in \Delta^n \right\} \subset\mathbb{R}^d.
\label{eqn:DeltamPhidefinition}
\end{align}
$\Phi \Delta^n$ is the convex hull of $P^1, \dots, P^n$ and is a bounded convex closed set in $\mathbb{R}^d$.
\subsection{Function $J(\bm{\lambda})$}
A function $J(\bm{\lambda})$, which serves as an analogue to the mutual information $I(\bm{\lambda})$, is defined \cite{sch} as follows. Let $\bm{u}\triangleq{^t}\left(\|P^1\|^2,\dots,\|P^n\|^2\right)\in\mathbb{R}^n$. Then, $J(\bm{\lambda})$ is defined by
\begin{align}
J(\bm{\lambda})&\triangleq(\bm{u},\bm{\lambda})-\|\Phi\bm{\lambda}\|^2\\
&=\sum_{i=1}^n\left(\sum_{j=1}^d(P^i_j)^2\right)\lambda_i-\sum_{j=1}^d\left(\sum_{i=1}^n P^i_j\lambda_i\right)^2,\,\bm{\lambda}\in\Delta^n.
\end{align}
By Jensen's inequality, we see that $J(\bm{\lambda})$ is a non-negative and concave function. As an analogue to (\ref{eqn:Cdefinition}), the following theorem holds.
\begin{theorem}
\label{the:Gamma=maxJlambda}{\rm (See \cite{sch})}
\begin{align}
\Gamma=\max_{\bm{\lambda}\in\Delta^n}J(\bm{\lambda}).
\end{align}
\end{theorem}
\noindent{\bf Proof:} A proof based on \cite{csi}, which differs from the proof in \cite{sch}, is presented in \ref{app:A}.$\hfill\square$
\section{On Equidistant Point}

We have the following theorem.

\begin{theorem}
\label{the:barycentriccoordinate}
Assume the rank condition (\ref{eqn:rankdefinition}). There exist a unique equidistant point $\widetilde{Q}$ from $P^1,\dots,P^n$, and a unique $\widetilde{\bm\lambda}\in L^n$ such that $\widetilde{Q}=\Phi\widetilde{\bm\lambda}\in\Phi L^n$.
\end{theorem}

\noindent{\bf Proof:} First, assume that there exist an equidistant point $\widetilde{Q}=\Phi\widetilde{\bm\lambda}\in\Phi L^n$ from $P^1,\dots,P^n$, with $\widetilde{Q}={^t}(\widetilde{Q}_1,\dots,\widetilde{Q}_d)$ and the barycentric coordinate $\widetilde{\bm\lambda}$ of $\widetilde{Q}$. We show that these $\widetilde{Q}$ and $\widetilde{\bm\lambda}$ are unique. The following notation will be used.
\begin{align}
P^{i+}&\triangleq{^t}(1,P^i_1,\cdots,P^i_d)=\begin{pmatrix}1\\ P^i\end{pmatrix}\in\mathbb{R}^{d+1},\,i=1,\cdots,n,\label{eqn:hatP}\\
\widetilde{Q}^+&\triangleq{^t}(1,\widetilde{Q}_1,\cdots,\widetilde{Q}_d)=\begin{pmatrix}1\\\widetilde{Q}\end{pmatrix}\in\mathbb{R}^{d+1},\label{eqn:hatQ}\\
\Phi^+&\triangleq(P^{1+},\dots,P^{n+})=\begin{pmatrix}{^t}\bm1\\\Phi\end{pmatrix}\in\mathbb{R}^{(d+1)\times n},\label{eqn:tbm1Phi}\\
\bm{1}&\triangleq{^t}(1,\cdots,1)\in\mathbb{R}^n,\\
H&\triangleq\begin{pmatrix}
-1 & 1 & 0 & \cdots & 0\\
-1 & 0 & 1 & \cdots & 0\\
\vdots & \vdots & \vdots & \ddots & \vdots\\
-1 & 0 & 0 & \cdots & 1\\ 
\end{pmatrix}\in\mathbb{R}^{(n-1)\times n},\label{eqn:Hdefinition}\\
\bm{u}&\triangleq{^t}(\|P^1\|^2,\dots,\|P^n\|^2)\in\mathbb{R}^n.\label{eqn:udefinition}
\end{align}

Under the rank condition (\ref{eqn:rankdefinition}), we have rank $\Phi^+=n$. This is because if we assume that $\sum_{i=1}^n\gamma_iP^{i+}=\begin{pmatrix}\sum_{i=1}^n\gamma_i\\\sum_{i=1}^n\gamma_iP^i\end{pmatrix}=\bm{0}$, then by calculation, we have $\sum_{i=2}^n\gamma_i(P^i-P^1)=\bm{0}$, hence by the rank condition (\ref{eqn:rankdefinition}), it follows that $\gamma_i=0$ for $i=1,\dots,n$.

Now, $\widetilde{Q}=\Phi\widetilde{\bm\lambda}$ and ${^t}\bm{1}\widetilde{\bm\lambda}=1$ together can be written as
\begin{align}
\widetilde{Q}^+=\Phi^+\widetilde{\bm\lambda}.\label{eqn:hatQ0}
\end{align}
Because $\widetilde{Q}$ is the equidistant point from $P^1,\dots,P^n$, it holds that $\|P^i-\widetilde{Q}\|^2=\|P^1-\widetilde{Q}\|^2,\,i=2,\dots,n$. Hence, by a simple calculation, we have
\begin{align}
2\,{^t}(P^i-P^1)\widetilde{Q}=\|P^i\|^2-\|P^1\|^2,\,i=2,\dots,n,\label{eqn:toukyori6}
\end{align}
thus, from (\ref{eqn:hatP}),(\ref{eqn:hatQ}),(\ref{eqn:toukyori6}),
\begin{align}
2\,{^t}(P^{i+}-P^{1+})\widetilde{Q}^+=\|P^{i+}\|^2-\|P^{1+}\|^2,\,i=2,\dots,n.\label{eqn:toukyori7}
\end{align}
If we write (\ref{eqn:toukyori7}) by using $H$ of (\ref{eqn:Hdefinition}), we have
\begin{align}
2H\,{^t}(\Phi^+)\widetilde{Q}^+=H\bm{u},\label{eqn:toukyori8}
\end{align}
then substitute (\ref{eqn:hatQ0}) into (\ref{eqn:toukyori8}), we have
\begin{align}
2H\,{^t}(\Phi^+)\Phi^+\widetilde{\bm\lambda}=H\bm{u}.\label{eqn:toukyori9}
\end{align}
Putting $M\triangleq2\,{^t}(\Phi^+)\Phi^+\in\mathbb{R}^{n\times n}$, we have $\text{rank}\,M=\text{rank}\,\Phi^+=n$ (\cite[p.13,\,0.4.6(d)]{hor}), then $M$ is a regular matrix. From (\ref{eqn:toukyori9}), we have
\begin{align}
H(M\widetilde{\bm\lambda}-\bm{u})=\bm{0}.\label{eqn:toukyori10}
\end{align}
Since Ker\,$H=\{\tau\,\bm{1}\,|\,\tau\in\mathbb{R}\}$, by (\ref{eqn:toukyori10}) there uniquely exists $\widetilde\tau\in\mathbb{R}$ such that
\begin{align}
\widetilde{\bm\lambda}=M^{-1}(\bm{u}+\widetilde\tau\bm{1}).\label{eqn:barycentric10}
\end{align}
By (\ref{eqn:barycentric10}) and ${^t}\bm{1}\widetilde{\bm\lambda}=1$, we have
\begin{align}
\widetilde\tau=\ds\frac{1-{^t}\bm{1}M^{-1}\bm{u}}{{^t}\bm{1}M^{-1}\bm{1}}.\label{eqn:tau1}
\end{align}
(Because $M$ is positive definite, ${^t}\bm{1}M^{-1}\bm{1}>0$. See Lemma \ref{lem:XtXispositivedefinite} below.) Therefore, $\widetilde{Q}=\Phi\widetilde{\bm\lambda}$ is unique (if exists).

Next, we show the existence of the equidistant point $\widetilde{Q}=\Phi\widetilde{\bm\lambda}$ from $P^1,\dots,P^n$. Define $\widetilde{\bm\lambda}$ using (\ref{eqn:barycentric10}) and (\ref{eqn:tau1}). By defining $\widetilde{Q} = \Phi \widetilde{\bm\lambda}$ with this $\widetilde{\bm\lambda}$, it can be seen that $\widetilde{Q} = \Phi \widetilde{\bm\lambda}$ is an equidistant point from $P^1,\dots,P^n$ by retracing the above discussion in reverse.$\hfill\square$

\medskip

The following lemma holds.

\begin{lemma}  
\label{lem:sumlambdaiPi-Q2}
For any $\bm\lambda\in\Delta^n$ and $Q\in\mathbb R^d$, we have
\begin{align}
\sum_{i=1}^n\lambda_i\|P^i-Q\|^2=J(\bm\lambda)+\|\Phi\bm\lambda-Q\|^2.\label{eqn:sumlambdaiPi-Q2}
\end{align}
\end{lemma}

\noindent{\bf Proof:} This follows from a straightforward calculation. $\hfill\square$

\bigskip

For the equidistant point $\widetilde{Q}$, the following lemma holds.

\begin{lemma}
\label{lem:maxJisminlambdaPhi-Q}
Assume the rank condition (\ref{eqn:rankdefinition}). The $\bm\lambda$ that achieves $\max_{\bm\lambda\in\Delta^n}J(\bm\lambda)$ coincides with the $\bm\lambda$ that achieves $\min_{\bm\lambda\in\Delta^n}\|\Phi\bm\lambda-\widetilde{Q}\|^2$.
\end{lemma}  
\noindent{\bf Proof:} Substituting $Q=\widetilde{Q}$ into (\ref{eqn:sumlambdaiPi-Q2}) and defining $\widetilde\Gamma\triangleq\|P^i-\widetilde{Q}\|^2,\,i=1,\dots,n$, we obtain $\widetilde\Gamma=J(\bm\lambda)+\|\Phi\bm\lambda-\widetilde{Q}\|^2$, which proves the lemma. $\hfill\square$

\bigskip

From Theorem \ref{the:Gamma=maxJlambda}, the barycentric coordinate $\bm\lambda^\ast$ of the center $Q^\ast$ of the SEB is given by the solution $\bm\lambda=\bm\lambda^\ast$ of the maximization problem $\max_{\bm\lambda\in\Delta^n}J(\bm\lambda)$. Combining this with Lemma \ref{lem:maxJisminlambdaPhi-Q}, we obtain the following theorem.
\begin{theorem}
\label{the:minlambdaPhitildeQ}
Assume the rank condition (\ref{eqn:rankdefinition}). The barycentric coordinate $\bm\lambda^\ast$ of the center $Q^\ast\in\Phi\Delta^n$ of the SEB is given by the solution $\bm\lambda=\bm\lambda^\ast$ of the minimization problem
\begin{align}
\min_{\bm\lambda\in\Delta^n}\|\Phi\bm\lambda-\widetilde{Q}\|^2.\label{eqn:minlambdaPhitildeQ}
\end{align}
$\hfill\square$
\end{theorem}

The objective function in (\ref{eqn:minlambdaPhitildeQ}) is represented by components as

\begin{align}  
\|\Phi\bm\lambda-\widetilde{Q}\|^2=\sum_{j=1}^d\left(\sum_{i=1}^nP^i_j\lambda_i-\widetilde{Q}_j\right)^2.\label{eqn:lambdaPhi-tildeQ2=lambdaiPij-Qj2}
\end{align}
\section{Two-Variable Formulation of Minimization Problem}
Eq.\,(\ref{eqn:minlambdaPhitildeQ}) is a one-variable minimization problem with $\bm\lambda \in \Delta^n$. Here, in order to examine (\ref{eqn:minlambdaPhitildeQ}), we consider a minimization problem
\begin{align}
\min_{\bm\lambda\in(L^n)'}\|\Phi\bm\lambda-\widetilde{Q}\|^2,\label{eqn:minlambdaLnPhilambda-Q}
\end{align}
where $\Delta^n$ is replaced with an arbitrary subset $(L^n)'$ of $L^n$, and then we construct a two-variable formulation for (\ref{eqn:minlambdaLnPhilambda-Q}). To do so, we first define, for any $Q' = {^t}(Q'_1, \dots, Q'_d) \in \mathbb{R}^d$,
\begin{align}
L^{n,d}(Q') \triangleq \left\{\Theta = (\theta_i^j)_{i=1,\dots,n,\,j=1,\dots,d} \in \mathbb{R}^{n \times d} \,\middle|\,\sum_{i=1}^n \theta_i^j = Q'_j,\, j = 1, \dots, d \right\}.
\end{align}
$\Theta$ corresponds to the backward channel in the Arimoto-Blahut algorithm.

We have the following theorem.
\begin{theorem}
\label{the:2hensuka}
For any subset $\left(L^n\right)'$ of $L^n$ and any $Q'={^t}(Q'_1,\dots,Q'_d) \in \mathbb{R}^d$, we have
\begin{align}
\ds \min_{\bm\lambda \in \left(L^n\right)'} \sum_{j=1}^d \left( \sum_{i=1}^n P^i_j \lambda_i - Q'_j \right)^2 = n \min_{\bm\lambda \in \left(L^n\right)'} \min_{\Theta \in L^{n,d}(Q')} \sum_{j=1}^d \sum_{i=1}^n \left( P^i_j \lambda_i - \theta_i^j \right)^2.\label{eqn:2hensuka}
\end{align}
\end{theorem}
\noindent {\bf Proof:} First, we prove that in (\ref{eqn:2hensuka}) the LHS (left-hand side) is less than or equal to the RHS (right-hand side). By the Cauchy-Schwarz inequality, for any real numbers $t_1, \dots, t_n$, we have $\left( \sum_{i=1}^n t_i \right)^2 \leq n \sum_{i=1}^n t_i^2$. Therefore,
\begin{align}
\left(\sum_{i=1}^nP^i_j\lambda_i-Q'_j\right)^2&=\left\{\sum_{i=1}^n\left(P^i_j\lambda_i-\theta_i^j\right)\right\}^2\\
&\leq n\sum_{i=1}^n\left(P^i_j\lambda_i-\theta_i^j\right)^2.
\end{align}
Summing over $j$, we have
\begin{align}
\sum_{j=1}^d\left(\sum_{i=1}^nP^i_j\lambda_i-Q'_j\right)^2\leq n\sum_{j=1}^d\sum_{i=1}^n\left(P^i_j\lambda_i-\theta_i^j\right)^2.
\end{align}
Taking the minimum on both sides, we obtain the inequality LHS $\leq$ RHS in (\ref{eqn:2hensuka}).

Next, we show that LHS $\geq$ RHS in (\ref{eqn:2hensuka}). Let $\bm\lambda=\widehat{\bm\lambda}$ be the minimizer of the LHS of (\ref{eqn:2hensuka}). For this $\widehat{\bm\lambda}$, define
\begin{align}
\widehat{\Theta}=(\widehat{\theta}_i^j),\,\widehat{\theta}_i^j\triangleq P^i_j\widehat{\lambda}_i-\dfrac{1}{n}\left(\sum_{k=1}^nP^k_j\widehat{\lambda}_k-Q'_j\right).\label{eqn:thetahatdefinition}
\end{align}
(Refer to (\ref{eqn:fromlambdatotheta}) for this choice of $\widehat{\theta}_i^j$). Summing over $i$ in (\ref{eqn:thetahatdefinition}), we have $\sum_{i=1}^n\widehat{\theta}_i^j=Q'_j$, thus $\widehat{\Theta}\in L^{n,d}(Q')$. For these $\widehat{\bm\lambda}$ and $\widehat{\Theta}$, from (\ref{eqn:thetahatdefinition}), we have
\begin{align}
n\sum_{j=1}^d\sum_{i=1}^n\left(P^i_j\widehat{\lambda}_i-\widehat{\theta}_i^j\right)^2&=n\sum_{j=1}^d\sum_{i=1}^n\dfrac{1}{n^2}\left(\sum_{k=1}^nP^k_j\widehat{\lambda}_k-Q'_j\right)^2\label{eqn:lambdaastThetaast}\\
&=\sum_{j=1}^d\left(\sum_{k=1}^nP^k_j\widehat{\lambda}_k-Q'_j\right)^2\\
&=\text{LHS}\text{ of (\ref{eqn:2hensuka})}.
\end{align}
Taking the minimum on the left-hand side of (\ref{eqn:lambdaastThetaast}), we obtain the inequality LHS $\geq$ RHS in (\ref{eqn:2hensuka}). \hfill $\square$

\subsection{Constructing Recurrence Formula via Alternating Minimization}

Now, in Theorem \ref{the:2hensuka}, let us consider $\Delta^n$ as a subset $(L^n)'$ of $L^n$ and consider $\widetilde{Q}$ as $Q'$. Based on Theorem \ref{the:2hensuka}, we aim to solve a two-variable minimization problem
\begin{align}
\min_{\bm\lambda\in\Delta^n}\min_{\Theta\in L^{n,d}(\widetilde{Q})}\sum_{j=1}^d\sum_{i=1}^n\left(P^i_j\lambda_i-\theta_i^j\right)^2.
\end{align}
However, minimizing with respect to $\bm\lambda\in\Delta^n$, i.e., $\lambda_i\geq0$ for $i=1,\dots,n$, is challenging due to the inequality constraints. Therefore, we first replace the minimization over $\bm\lambda\in\Delta^n$ with the minimization over $\bm\lambda\in L^n$. Then, we solve the minimization problem over $\bm\lambda\in L^n$ and $\Theta\in L^{n,d}(\widetilde{Q})$ to derive a recurrence formula for $\bm\lambda$, which we then modify to obtain a recurrence formula for $\bm\lambda\in\Delta^n$.

In the following, we solve the minimization problem
\begin{align}
\min_{\bm\lambda\in L^n}\min_{\Theta\in L^{n,d}(\widetilde{Q})}\sum_{j=1}^d\sum_{i=1}^n\left(P^i_j\lambda_i-\theta_i^j\right)^2\label{eqn:minlambdaLmminThetaLm}
\end{align}
by employing an alternating minimization algorithm. Specifically, we iteratively minimize with respect to one variable while keeping the other fixed.

\subsubsection{Minimizing $\Theta$ given $\bm\lambda$}
For a fixed $\bm\lambda \in L^n$, we minimize the objective function $\sum_{j=1}^d\sum_{i=1}^n(P^i_j\lambda_i-\theta_i^j)^2$ in (\ref{eqn:minlambdaLmminThetaLm}) with respect to $\Theta\in L^{n,d}(\widetilde{Q})$. Using Lagrange multipliers $\ell_1, \dots, \ell_d$, we define
\begin{align}
f(\Theta)\triangleq\sum_{j'=1}^d\sum_{i'=1}^n\left(P^{i'}_{j'}\lambda_{i'}-\theta_{i'}^{j'}\right)^2+\sum_{j'=1}^d\ell_{j'}\left(\sum_{i'=1}^n\theta_{i'}^{j'}-\widetilde{Q}_{j'}\right).
\end{align}
By solving $\partial f(\Theta)/\partial\theta_i^j=0$ with $\sum_{i=1}^n \theta_i^j = \widetilde{Q}_j$, the optimal $\theta_i^j$ is given by
\begin{align}
\theta_i^j=P^i_j\lambda_i-\dfrac{1}{n}(Q_j-\widetilde{Q}_j),\ i=1,\dots,n,j=1,\dots,d,\label{eqn:fromlambdatotheta}
\end{align}
where $Q_j\triangleq\sum_{i=1}^nP^i_j\lambda_i$.
\subsubsection{Minimizing $\bm\lambda$ given $\Theta$}
Next, for a fixed $\Theta \in L^{n,d}(\widetilde{Q})$, we minimize $\sum_{j=1}^d\sum_{i=1}^n(P^i_j\lambda_i-\theta_i^j)^2$ in (\ref{eqn:minlambdaLmminThetaLm}) with respect to $\bm\lambda\in L^n$. Using Lagrange multiplier $\ell$, we define
\begin{align}
g(\bm\lambda)\triangleq\sum_{j=1}^d\sum_{i'=1}^n\left(P^{i'}_j\lambda_{i'}-\theta_{i'}^j\right)^2+\ell\left(\sum_{i'=1}^n\lambda_{i'}-1\right).
\end{align}
By solving $\partial g(\bm\lambda)/\partial\lambda_i=0$ with $\sum_{i=1}^n\lambda_i=1$, the optimal $\lambda_i$ is given by
\begin{align}
\lambda_i=\dfrac{1}{\|P^i\|^2}\left\{a_i-\dfrac{1}{s}\left(\sum_{k=1}^n\dfrac{a_{k}}{\|P^{k}\|^2}-1\right)\right\},\ i=1,\dots,n,\label{eqn:fromthetatolambda}
\end{align}
where
\begin{align}
a_i\triangleq\ds\sum_{j=1}^d\theta_i^jP^i_j,\label{eqn:aidefinition}
\end{align}
\begin{align}
s\triangleq\sum_{i=1}^n\dfrac{1}{\|P^i\|^2}.\label{eqn:sdefinition}
\end{align}

By substituting (\ref{eqn:fromlambdatotheta}) into (\ref{eqn:fromthetatolambda})-(\ref{eqn:sdefinition}), a recurrence formula for $\bm\lambda^N$ is obtained. Consider $\bm\lambda$ of (\ref{eqn:fromlambdatotheta}) as $\bm\lambda^N$ and $\bm\lambda$ of (\ref{eqn:fromthetatolambda}) as $\bm\lambda^{N+1}$, then we have the recurrence formula as follows.

First, substituting (\ref{eqn:fromlambdatotheta}) into (\ref{eqn:aidefinition}), we have
\begin{align}
a_i=\lambda_i\|P^i\|^2-\dfrac{1}{n}\left(P^i,Q-\widetilde{Q}\right),\ i=1,\dots,n.\label{eqn:ai=lambdaiPi2-1nPiQ-Q}
\end{align}
Then next, substituting (\ref{eqn:ai=lambdaiPi2-1nPiQ-Q}) into (\ref{eqn:fromthetatolambda}) with $Q^N=\Phi\bm\lambda^N$, we have
\begin{align}
\lambda^{N+1}_i=\lambda^N_i-\dfrac{1}{n\|P^i\|^2}\left(P^i,Q^N-\widetilde{Q}\right)+\dfrac{1}{ns\|P^i\|^2}\sum_{k=1}^n\dfrac{\left(P^{k},Q^N-\widetilde{Q}\right)}{\|P^{k}\|^2},\ i=1,\dots,n.\label{eqn:lambdarecurrence}
\end{align}
\subsection{Recurrence Formula without using Equidistant Point $\widetilde{Q}$}

We made a recurrence formula for $\bm\lambda^N$ by (\ref{eqn:lambdarecurrence}) for the time being, but (\ref{eqn:lambdarecurrence}) still contains $\widetilde{Q}$, which we cannot obtain by an easy calculation. Therefore, we replace the formula (\ref{eqn:lambdarecurrence}) with one that does not contain $\widetilde{Q}$, using only quantities that can be easily computed from $\Phi$. This substitution allows the recurrence formula to be applicable to any $\Phi$ without assuming the existence of $\widetilde{Q}$.

From (\ref{eqn:lambdarecurrence}), we have
\begin{align}
\lambda^{N+1}_i&=\lambda^N_i+\dfrac{1}{ns\|P^i\|^2}\left\{\sum_{k=1}^n\dfrac{(P^{k},Q^N-\widetilde{Q})}{\|P^{k}\|^2}-s(P^i,Q^N-\widetilde{Q})\right\}\\[1mm]
&=\lambda^N_i+\dfrac{1}{ns\|P^i\|^2}\sum_{k=1}^n\dfrac{(P^{k}-P^i,Q^N-\widetilde{Q})}{\|P^{k}\|^2}\\[1mm]
&=\lambda^N_i+\underline{\dfrac{1}{ns\|P^i\|^2}\sum_{k=1}^n\dfrac{(P^{k}-P^i,Q^N)}{\|P^{k}\|^2}}_{\,\star1}+\underline{\left\{-\dfrac{1}{ns\|P^i\|^2}\sum_{k=1}^n\dfrac{(P^{k}-P^i,\widetilde{Q})}{\|P^{k}\|^2}\right\},}_{\,\star2}\label{eqn:lambdaN+1=lmabdaN+alpha}\\
&i=1,\dots,n.\nonumber
\end{align}

Now, we calculate the two terms $\star 1$ and $\star 2$ in (\ref{eqn:lambdaN+1=lmabdaN+alpha}).
\subsection{Calculation of $\star 1$ in (\ref{eqn:lambdaN+1=lmabdaN+alpha})}

First, we calculate $\star 1$ in (\ref{eqn:lambdaN+1=lmabdaN+alpha}). We note that $Q^N = \Phi \bm\lambda^N$, so
\begin{align}
\star1&=\dfrac{1}{ns\|P^i\|^2}\left\{\sum_{k=1}^n\dfrac{1}{\|P^{k}\|^2}(P^{k}-P^i),\Phi\bm\lambda^N\right\}\\
&=\sum_{i'=1}^n\lambda^N_{i'}\times\dfrac{1}{n\|P^i\|^2}\left(\dfrac{1}{s}\sum_{k=1}^n\dfrac{1}{\|P^k\|^2}P^k-P^i,P^{i'}\right)\label{eqn:1/SPkPk2PiPi'}\\
&=\sum_{i'=1}^n\lambda^N_{i'}\times\dfrac{\left(\overline{P}-P^i,P^{i'}\right)}{n\|P^i\|^2},\label{eqn:barP-PiPi'}
\end{align}
where
\begin{align}
\overline{P}\triangleq\dfrac{1}{s}\sum_{k=1}^n\dfrac{1}{\|P^k\|^2}P^k\in\Phi\Delta^n.\label{eqn:barPdefinition}
\end{align}

For calculating (\ref{eqn:barP-PiPi'}), we define
\begin{align}
B\triangleq(b_{ii'})_{i,i'=1,\dots,n}\in{\mathbb R}^{n\times n},\ b_{ii'}=\dfrac{\left(\overline{P}-P^i,P^{i'}\right)}{n\|P^i\|^2},\label{eqn:bi'idefinition}
\end{align}
\begin{align}
T^i\triangleq\dfrac{1}{n\|P^i\|^2}(\overline{P}-P^i)\in{\mathbb R}^d,\ i=1,\dots,n,\label{eqn:Tidefinition}
\end{align}
and
\begin{align}
T\triangleq\begin{pmatrix}T^1,\dots,T^n\end{pmatrix}\in\mathbb{R}^{d\times n}.\label{eqn:Psidefinition}
\end{align}

From (\ref{eqn:bi'idefinition})-(\ref{eqn:Psidefinition}), we have $b_{ii'}=(T^i,P^{i'})={^t}T^iP^{i'}$, thus $B={^t}T\Phi\in\mathbb R^{n\times n}$. Then we can calculate $T$ as follows. (\ref{eqn:barPdefinition}) becomes
\begin{align}
\overline{P}&=\dfrac{1}{s}(P^1,\dots,P^n)\,\text{diag}\left(\dfrac{1}{\|P^1\|^2},\dots,\dfrac{1}{\|P^n\|^2}\right)\begin{pmatrix}1\\\vdots\\1\end{pmatrix}\\
&=\dfrac{1}{s}\Phi D\bm1,\label{eqn:dfrac1s1DPhi}
\end{align}
where we put
\begin{align}
D\triangleq\text{diag}\left(\dfrac{1}{\|P^1\|^2},\dots,\dfrac{1}{\|P^n\|^2}\right)\in\mathbb R^{n\times n},\label{eqn:Ddefinition}
\end{align}
which is the diagonal matrix with diagonal components $1/\|P^1\|^2,\dots,1/\|P^n\|^2$. Next, from (\ref{eqn:Tidefinition}),(\ref{eqn:Psidefinition}),
\begin{align}
T&=(T^1,\dots,T^n)\\
&=\left(\dfrac{1}{n\|P^1\|^2}(\overline{P}-P^1),\dots,\dfrac{1}{n\|P^n\|^2}(\overline{P}-P^n)\right)\\
&=\dfrac{1}{n}\left(\overline{P}-P^1,\dots,\overline{P}-P^n\right)\,\text{diag}\left(\dfrac{1}{\|P^1\|^2},\dots,\dfrac{1}{\|P^n\|^2}\right)\\
&=\dfrac{1}{n}(\overline{P}\,{^t}\bm1-\Phi)D\\
&=\dfrac{1}{ns}\Phi(D\bm1{^t}\bm1-sI)D\ \ (\text{from }(\ref{eqn:dfrac1s1DPhi})),\label{eqn:Trepresentation}
\end{align}
where $I\in\mathbb R^{n\times n}$ is the $n$-dimensional identity matrix. Noting that $D={^t}D$, we have
\begin{align}
B&={^t}T\Phi\\
&=\dfrac{1}{ns}D\left(\bm1{^t}\bm1D-sI\right){^t}\Phi\Phi\ \ (\text{from }(\ref{eqn:Trepresentation}))\\
&=-\dfrac{1}{ns}\left(sD-D\bm1{^t}\bm1D\right){^t}\Phi\Phi\\
&=-\Omega\,{^t}\Phi\Phi,\label{eqn:PhitPhiOmegais-B}
\end{align}
where
\begin{align}
\Omega\triangleq\dfrac{1}{ns}\left(sD-D\bm1{^t}\bm1D\right).\label{eqn:Omegadefinition2}
\end{align}
\subsection{Calculation of $\star 2$ in (\ref{eqn:lambdaN+1=lmabdaN+alpha})}

Next, we calculate $\star 2$ in (\ref{eqn:lambdaN+1=lmabdaN+alpha}). Since $\widetilde{Q}$ is the equidistant point from $P^1,\dots,P^n$, we have $\|P^k-\widetilde{Q}\|^2=\|P^i-\widetilde{Q}\|^2$ for any $k,i$. By a simple calculation, we find that $(P^k - P^i, \widetilde{Q}) = \frac{1}{2}(\|P^k\|^2 - \|P^i\|^2)$, which allows us to eliminate $\widetilde{Q}$. Therefore, we have
\begin{align}
\star2&=-\dfrac{1}{2}\dfrac{1}{ns\|P^i\|^2}\sum_{k=1}^n\dfrac{\|P^k\|^2-\|P^i\|^2}{\|P^k\|^2}\\[1mm]
&=-\dfrac{1}{2}\dfrac{1}{ns\|P^i\|^2}(n-s\|P^i\|^2)\\
&=\dfrac{1}{2n}-\dfrac{1}{2s\|P^i\|^2},\ i=1,\dots,n.\label{eqn:12m-12sPi2}
\end{align}
We write (\ref{eqn:12m-12sPi2}) in vector form. Let
\begin{align}
\bm{c}={^t}(c_1,\dots,c_n),\ c_i\triangleq\dfrac{1}{2n}-\dfrac{1}{2s\|P^i\|^2},\label{eqn:cdefinition}
\end{align}
then
\begin{align}
&\bm{c}=\dfrac{1}{2n}\bm{1}-\dfrac{1}{2s}D\bm1.\label{eqn:vectorcdefinition}
\end{align}
From (\ref{eqn:lambdaN+1=lmabdaN+alpha}),(\ref{eqn:PhitPhiOmegais-B}),(\ref{eqn:vectorcdefinition}), we have $\bm\lambda^{N+1}=\bm\lambda^N+B\bm\lambda^N+\bm{c}=(I-\Omega\,{^t}\Phi\Phi)\bm\lambda^N+\bm{c}$. 

Summing up the above discussion, we obtain the following recurrence formula.

\subsection{Recurrence Formula for $\bm\lambda^N$}
\label{sec:RecurrenceFormulaforlambdaN}
\begin{align}
\bm\lambda^{N+1}&=R\bm\lambda^N+\bm{c}\in\mathbb R^n,\label{eqn:proposed1}\\
R&=I-\Omega\,{^t}\Phi\Phi\in\mathbb R^{n\times n},\label{eqn:proposed2}\\
\bm{c}&=\dfrac{1}{2n}\bm{1}-\dfrac{1}{2s}D\bm1\in\mathbb R^n,\label{eqn:proposed3}\\
\Phi&=(P^1,\dots,P^n)\in\mathbb R^{d\times n},\label{eqn:proposed4}\\
\Omega&=\dfrac{1}{ns}(sD-D\bm1{^t}\bm1D)\in\mathbb R^{n\times n},\label{eqn:proposed5}\\
D&=\text{diag}\left(\dfrac{1}{\|P^1\|^2},\dots,\dfrac{1}{\|P^n\|^2}\right)\in\mathbb R^{n\times n},\label{eqn:proposed6}\\
s&=\sum_{i=1}^n\dfrac{1}{\|P^i\|^2}=\text{tr}\,D={^t}\bm1D\bm1\in\mathbb R,\label{eqn:proposed7}\\
\bm1&={^t}(1,\dots,1)\in\mathbb R^n.\label{eqn:proposed8}
\end{align}
\begin{remark}
\rm This recurrence formula uses only very simple terms, so it is easy to make a computation program. Further, as a result, the recurrence formula became independent of the rank condition (\ref{eqn:rankdefinition}) and the existence of the equidistant point $\widetilde{Q}$. The recurrence formula can be computed using only $\Phi$, so it is applicable to any $\Phi$. When the rank condition (\ref{eqn:rankdefinition}) holds, $\bm{\lambda}^N$ converges to the barycentric coordinates $\widetilde{\bm{\lambda}}$ of the equidistant point $\widetilde{Q}$ as $N \to \infty$. See Theorem $\ref{the:mainsecond}$ below. If the rank condition does not hold, convergence will be evaluated numerically in Sections \ref{sec:heuristic} and \ref{sec:numericalexamples2}.
\end{remark}
\begin{remark}
\rm Since (\ref{eqn:proposed6}) and (\ref{eqn:proposed7}) include $1/\|P^i\|^2$, we need to ensure that $P^i\neq\bm{0}$. If there is any $P^i = \bm{0}$, the points $P^1, \dots, P^n$ should be translated simultaneously to ensure that $P^i \neq \bm{0}$ for all $i = 1, \dots, n$. To achieve it, for example, the dimensionality of the points can be increased by adding a constant, such as ${^t}(P^i_1,\dots,P^i_d)\to{^t}(1,P^i_1,\dots,P^i_d)$. Simultaneous translation does not affect the barycentric coordinates.
\end{remark}
\section{Analysis for Convergence of Recurrence Formula}
\label{sec:convergencespeed}
For the sequence $\{\bm{\lambda}^N\}_{N=0,1,\dots}$ defined by the recurrence formula (\ref{eqn:proposed1}), we prove, under the rank condition (\ref{eqn:rankdefinition}), that $\bm{\lambda}^N\to\widetilde{\bm{\lambda}}$ as $N\to\infty$, where $\widetilde{\bm{\lambda}}$ is the barycentric coordinate of the equidistant point $\widetilde{Q}$. To prove it, we will use the following lemmas.
\begin{lemma}
\label{lem:lN+1-tildel=lN-tildelR}
Assume the rank condition (\ref{eqn:rankdefinition}). We have
\begin{align}
\bm\lambda^{N+1}-\widetilde{\bm\lambda}=R(\bm\lambda^N-\widetilde{\bm\lambda}).\label{eqn:lN+1-tildel=lN-tildelR-2}
\end{align}
\end{lemma}
\noindent{\bf Proof:} First, from (\ref{eqn:Omegadefinition2}),(\ref{eqn:proposed7}), we have
\begin{align}
\Omega\bm1=\dfrac{1}{ns}(sD-D\bm1{^t}\bm1D)\bm1=\bm0.\label{eqn:Omega1=0}
\end{align}
Thus, from (\ref{eqn:tbm1Phi}),(\ref{eqn:Omega1=0}),
\begin{align}
\Omega\,{^t}(\Phi^+)\Phi^+&=\Omega(\bm1,{^t}\Phi)\begin{pmatrix}{^t}\bm1\\\Phi\end{pmatrix}\\
&=\Omega\bm1{^t}\bm1+\Omega\,{^t}\Phi\Phi\\
&=\Omega\,{^t}\Phi\Phi.\label{eqn:OmegaPhi+Phi+=OmegaPhiPhi}
\end{align}
From (\ref{eqn:proposed1}), we have
\begin{align}
\bm\lambda^{N+1}-\widetilde{\bm\lambda}=R(\bm\lambda^N-\widetilde{\bm\lambda})+\bm{c}-(I-R)\widetilde{\bm\lambda}.\label{eqn:lN+1-tildel=RlN-tildel+c-(I-R)tildel}
\end{align}
Then, we calculate $(I-R)\widetilde{\bm{\lambda}}$ to obtain
\begin{align}
(I-R)\widetilde{\bm\lambda}&=\Omega\,{^t}\Phi\Phi\widetilde{\bm\lambda}\ \ ({\rm from}\ (\ref{eqn:proposed2}))\\
&=\Omega\,{^t}(\Phi^+)\Phi^+\widetilde{\bm\lambda}\ \ ({\rm from} \ (\ref{eqn:OmegaPhi+Phi+=OmegaPhiPhi}))\\
&=\Omega\,\dfrac{1}{2}MM^{-1}(\bm{u}+\widehat\tau\bm1)\ \ ({\rm from}\ (\ref{eqn:barycentric10}))\\
&=\dfrac{1}{2}\Omega\bm{u}\ \ ({\rm from}\ (\ref{eqn:Omega1=0}))\\
&=\dfrac{1}{2}\left(\dfrac{1}{n}D\bm{u}-\dfrac{1}{ns}D\bm1{^t}\bm1D\bm{u}\right)\ \ ({\rm from}\ (\ref{eqn:proposed5}))\\
&=\dfrac{1}{2n}\bm1-\dfrac{1}{2s}D\bm1\ \ ({\rm from}\ (\ref{eqn:udefinition}),(\ref{eqn:proposed6}))\\
&=\bm{c}\ \ (\text{from }(\ref{eqn:proposed3})).\label{eqn:(I-R)lambda=c}
\end{align}

Thus, from (\ref{eqn:lN+1-tildel=RlN-tildel+c-(I-R)tildel}),(\ref{eqn:(I-R)lambda=c}), the lemma is proved. \hfill$\square$

\bigskip

Now, let $\bm{\mu}^N\triangleq\bm{\lambda}^N-\widetilde{\bm{\lambda}}$, then from Lemma \ref{lem:lN+1-tildel=lN-tildelR}, we have
\begin{align}
\bm\mu^{N+1}=R\bm\mu^N,\ N=0,1,\dots.\label{eqn:muN+1=muNR}
\end{align}
We represent $\bm{\mu}^N$ by components as $\bm{\mu}^N={^t}(\mu_1^N,\dots,\mu_n^N)$. We study the convergence of the recurrence formula (\ref{eqn:muN+1=muNR}) as $N\to\infty$. For this, we will investigate the eigenvalues of $R$.
\subsection{Eigenvalues of $R$}
To analyze the eigenvalues of $R$, we prepare the following lemmas.
\subsubsection{Non-negative Definite and Positive Definite Matrices}
A matrix $X\in\mathbb{R}^{n \times n}$ is said to be {\it non-negative definite} if it is symmetric and ${^t}\bm{x}X\bm{x} \geq 0$ holds for all $\bm{x} \in \mathbb{R}^n$. $X$ being non-negative definite is represented by $X\geq O$. A matrix $X\in\mathbb{R}^{n\times n}$ is {\it positive definite} if it is symmetric and ${^t}\bm{x}X\bm{x}>0$ holds for all $\bm{x}\in\mathbb{R}^n,\,\bm{x}\neq\bm{0}$. $X$ being positive definite is represented by $X>O$. Every positive definite matrix is also non-negative definite.

The following lemma holds.

\begin{lemma}
\label{lem:Omegaisnonnegativedefinite}
$\Omega$ is non-negative definite.
\end{lemma}
\noindent{\bf Proof:} From (\ref{eqn:proposed5}), $\Omega$ is symmetric. We then show that ${^t}\bm{t}(sD-D\bm{1}{^t}\bm{1}D)\bm{t}\geq 0$ holds for any $\bm{t}={^t}(t_1,\dots,t_n)\in\mathbb{R}^n$. Let $D_i\triangleq 1/\|P^i\|^2,\,i=1,\dots,n$, then we have
\begin{align}
{^t}\bm{t}(sD-D\bm1{^t}\bm1D)\bm{t}&={^t}\bm1D\bm1{^t}\bm{t}D\bm{t}-{^t}\bm{t}D\bm1\,{^t}({^t}\bm{t}D\bm1)\ \ (\text{from}\ (\ref{eqn:proposed7}))\\
&=\sum_{i=1}^nD_i\sum_{i=1}^nD_it_i^2-\left(\sum_{i=1}^nD_it_i\right)^2\\
&\geq0\ \ (\text{by the Cauchy-Schwarz inequality}).\label{eqn:Omegaisnonnegativedefinite}
\end{align}
For equality to hold in (\ref{eqn:Omegaisnonnegativedefinite}), the necessary and sufficient condition is $\bm{t}=\sigma\bm{1},\,\sigma\in\mathbb{R}$. \hfill $\square$
\begin{lemma}
\label{lem:traceislessthanequal1}
We have $\text{\rm tr}\,\Omega\,{^t}\Phi\Phi\leq1$, where {\rm tr} denotes the trace of matrix.
\end{lemma}
\noindent{\bf Proof:} From (\ref{eqn:bi'idefinition}),(\ref{eqn:PhitPhiOmegais-B}), we have
\begin{align}
\text{tr}\,\Omega\,{^t}\Phi\Phi&=-\text{tr}\,B\\
&=-\sum_{i=1}^n\dfrac{\left(\overline{P}-P^i,P^i\right)}{n\|P^i\|^2}\\
&=-\dfrac{1}{n}\sum_{i=1}^n\dfrac{(\overline{P},P^i)-(P^i,P^i)}{\|P^i\|^2}\\
&=-\dfrac{1}{n}\left\{\left(\overline{P},\sum_{i=1}^n\dfrac{1}{\|P^i\|^2}P^i\right)-\sum_{i=1}^n\dfrac{\|P^i\|^2}{\|P^i\|^2}\right\}\\
&=-\dfrac{1}{n}\left\{\left(\overline{P},s\overline{P}\right)-n\right\}\\
&=1-\dfrac{s}{n}\|\overline{P}\|^2\\
&\leq1.
\end{align}
\hfill$\square$

We have the following theorem.
\begin{theorem}
\label{the:main}
Assume the rank condition (\ref{eqn:rankdefinition}). All eigenvalues of $R$ are real. Let the eigenvalues of $R$, in descending order, be $\eta_1, \eta_2, \dots, \eta_n$, then
\begin{align}
1=\eta_1>\eta_2\geq\dots\geq\eta_n\geq0.\label{eqn:1=theta1>theta2thetam0}
\end{align}
\end{theorem}

To prove this theorem, we use several lemmas.
\begin{lemma}
\label{lem:XtXispositivedefinite}
\text{\rm (\cite[Theorem 7.2.7(a), Theorem 7.1.7]{hor})} For any $X\in\mathbb{R}^{n \times m}$, $X\,{^t}X\in{\mathbb R}^{n\times n}$ is non-negative definite, i.e., $X\,{^t}X\geq O$. If $X\,{^t}X$ is regular, then it is positive definite, i.e., $X\,{^t}X>O$. \hfill $\square$
\end{lemma}
\begin{lemma}
\label{lem:squarerootofX}
\text{\rm(\cite[Theorem 7.2.6(a)]{hor})} For any non-negative definite matrix $X\in\mathbb{R}^{n \times n}$, there exists a non-negative definite matrix $Y\in\mathbb{R}^{n \times n}$ such that $X=Y^2$. Moreover, for any positive definite matrix $X\in\mathbb{R}^{n \times n}$, there exists a positive definite matrix $Y\in\mathbb{R}^{n \times n}$ such that $X=Y^2$.\hfill$\square$
\end{lemma}
\begin{lemma}
\label{lem:eigenvaluesofXY=YX}
\text{\rm(\cite[Theorem 1.3.22]{hor})} Let $X\in\mathbb{R}^{n \times m}$ and $Y\in\mathbb{R}^{m \times n}$, where $n\leq m$. Then, the $m$ eigenvalues of $YX\in\mathbb{R}^{m \times m}$ consist of the $n$ eigenvalues of $XY\in\mathbb{R}^{n \times n}$ along with $m-n$ additional zeros. That is, the characteristic polynomials of $YX$ and $XY$ satisfy the relation $\det(YX-tI)=t^{m-n}\det(XY-tI)$. In particular, when $n=m$, the $n$ eigenvalues of $XY\in\mathbb{R}^{n \times n}$ coincide with those of $YX\in\mathbb{R}^{n \times n}$. We denote by $XY\sim YX$ that the eigenvalues of $XY$ and $YX$ coincide.\hfill$\square$
\end{lemma}
\begin{lemma}
\label{lem:eigenvaluofndmisnn}
\text{\rm(\cite[Observation 7.1.4]{hor})} All eigenvalues of a non-negative definite matrix $X\in\mathbb{R}^{n \times n}$ are non-negative real numbers.\hfill$\square$
\end{lemma}

Now, when all eigenvalues of a matrix $X\in\mathbb{R}^{n \times n}$ are real, let $\nu_+(X),\,\nu_-(X),\,\nu_0(X)$ denote the number of positive, negative, and zero eigenvalues of $X$, respectively.

\smallskip

\underline{Definition of Inertia} (\cite[Definition 4.5.6]{hor}) Let $X\in\mathbb{R}^{n \times n}$ be a symmetric matrix. Then, all eigenvalues of $X$ are real. The ordered triple $(\nu_+(X),\,\nu_-(X),\,\nu_0(X))$ is called the {\it inertia} of $X$ and is denoted by $\text{In}(X)\triangleq(\nu_+(X),\,\nu_-(X),\,\nu_0(X))$.
%

\underline{Definition of Congruence} (\cite[Definition 4.5.4]{hor}) Two matrices $X,Y\in\mathbb{R}^{n \times n}$ are said to be {\it congruent} if there exists a regular matrix $Z\in\mathbb{R}^{n \times n}$ such that $Y=ZX\,{^t}Z$. Congruence is an equivalence relation (\cite[Theorem 4.5.8]{hor}). We denote the congruence of $X$ and $Y$ by $X\equiv Y$.
\begin{lemma}
\label{lem:Sylvester}
\text{\rm(Sylvester's Law of Inertia, \cite[Theorem 4.5.8]{hor})} For two symmetric matrices $X, Y\in\mathbb{R}^{n \times n}$, if $X\equiv Y$ then $\text{\rm In}(X)=\text{\rm In}(Y)$.\hfill$\square$
\end{lemma}

\bigskip

We prove Theorem \ref{the:main} using the above lemmas.

\noindent{\bf Proof of Theorem \ref{the:main}:} First, we show that all eigenvalues of $\Omega\,{^t}\Phi\Phi\in\mathbb R^{n\times n}$ are non-negative real numbers and satisfy
\begin{align}
\nu_+(\Omega\,{^t}\Phi\Phi)=n-1,\,\nu_-(\Omega\,{^t}\Phi\Phi)=0,\,\nu_0(\Omega\,{^t}\Phi\Phi)=1.\label{eqn:OmegaPhin+n-n0}
\end{align}
We have
\begin{align}
\Omega\,{^t}\Phi\Phi&=\Omega\,{^t}(\Phi^+)\Phi^+\ \ \text{(from (\ref{eqn:OmegaPhi+Phi+=OmegaPhiPhi}))}\label{eqn:PhitPhiOmegareference}\\
&=\Omega X^2\ \ (\exists X>O,\text{from Lemmas }\ref{lem:XtXispositivedefinite},\ref{lem:squarerootofX},\text{and the statement below }(\ref{eqn:toukyori9}))\\
&\sim X\Omega X\ \ \text{(from Lemma \ref{lem:eigenvaluesofXY=YX})}\\
&=X\Omega\,{^t}\!X\ \ (\text{because }X>O\text{ implies }X={^t}\!X)\\
&\equiv\Omega\ \ (\text{because }X>O\text{ ensures }X\text{ is regular}).\label{eqn:equivOmega}
\end{align}

From (\ref{eqn:equivOmega}) and Lemma \ref{lem:Sylvester}, we have $\text{In}(X\Omega\,{^t}X)=\text{In}(\Omega)$. By Lemmas \ref{lem:Omegaisnonnegativedefinite} and \ref{lem:eigenvaluofndmisnn}, all eigenvalues of $\Omega$ are non-negative real numbers. Therefore, from (\ref{eqn:PhitPhiOmegareference})-(\ref{eqn:equivOmega}), we have $\nu_+(\Omega\,{^t}\Phi\Phi)=\nu_+(\Omega),\,\nu_-(\Omega\,{^t}\Phi\Phi)=\nu_-(\Omega)=0,\,\nu_0(\Omega\,{^t}\Phi\Phi)=\nu_0(\Omega)$. Thus, to prove (\ref{eqn:OmegaPhin+n-n0}), it suffices to show  
\begin{align}
\nu_+(\Omega)=n-1,\,\nu_0(\Omega)=1.\label{eqn:n+(Omega)=m-1,n0(Omega)=1}
\end{align}
To prove it, we consider the null space $\mathcal{N}(\Omega)$ of the linear transformation $\Omega:\mathbb{R}^n\to\mathbb{R}^n$, defined as $\mathcal{N}(\Omega)\triangleq\{\bm{t}\in\mathbb{R}^n\mid\Omega\bm{t}=\bm{0}\}$. From (\ref{eqn:Omega1=0}), we have $\bm{1}\in\mathcal{N}(\Omega)$. Since any $\bm{t}\in\mathcal{N}(\Omega)$ satisfies ${^t}\bm{t}\,\Omega\,\bm{t}=0$, it follows from the proof of Lemma \ref{lem:Omegaisnonnegativedefinite} that $\bm{t}=\sigma\bm{1},\sigma\in\mathbb{R}$. Thus, $\mathcal{N}(\Omega)=\{\sigma\bm{1}\mid\sigma\in\mathbb{R}\}$, implying that $\dim\mathcal{N}(\Omega)=1$. By the rank-nullity theorem (\cite[Eq.[0.2.3.1]{hor}]), $\text{rank}(\Omega)+\dim\mathcal{N}(\Omega)=n$, which gives $\text{rank}(\Omega)=n-1$. Since $\text{rank}(\Omega)=\nu_+(\Omega)$ and $\nu_+(\Omega)+\nu_0(\Omega)=n$, we obtain (\ref{eqn:n+(Omega)=m-1,n0(Omega)=1}), proving (\ref{eqn:OmegaPhin+n-n0}).

Denoting the eigenvalues of $\Omega\,{^t}\Phi\Phi$ in ascending order as $\rho_1, \rho_2,\dots,\rho_n$, then by (\ref{eqn:OmegaPhin+n-n0}), we have $0=\rho_1<\rho_2\leq\dots\leq\rho_n$. Moreover, from Lemma \ref{lem:traceislessthanequal1}, $\rho_n\leq\sum_{i=1}^n\rho_i=\text{tr}(\Omega\,{^t}\Phi\Phi)\leq 1$. Thus, we obtain
\begin{align}
0=\rho_1<\rho_2\leq\dots\leq\rho_n\leq1.\label{eqn:0=beta1<beta2leqbetamleq1}
\end{align}
Since the eigenvalues $\eta_i$ of $R=I-\Omega\,{^t}\Phi\Phi$ satisfy $\eta_i=1-\rho_i$, it follows from (\ref{eqn:0=beta1<beta2leqbetamleq1}) that
\begin{align}
1=\eta_1>\eta_2\geq\dots\geq\eta_n\geq 0.\label{eqn:1=eta1>eta2geqetangeq0}
\end{align}
This completes the proof of Theorem \ref{the:main}.\hfill$\square$
\subsection{Convergence of Recurrence Formula (\ref{eqn:proposed1}) to $\widetilde{\bm\lambda}$}

We investigate the convergence of the recurrence formula (\ref{eqn:proposed1}) and (\ref{eqn:muN+1=muNR}) and their convergence rate.
\begin{theorem}
\label{the:mainsecond}
We assume the rank condition (\ref{eqn:rankdefinition}). The sequence $\{\bm\lambda^N\}_{N=0,1,\dots}$ defined by the recurrence formula (\ref{eqn:proposed1}) converges to $\widetilde{\bm\lambda}$, equivalently, the sequence $\{\bm\mu^N\}_{N=0,1,\dots}$ defined by (\ref{eqn:muN+1=muNR}) converges to $\bm{0}$, and the convergence rate is given by
\begin{align}
\|\bm\mu^N\|\leq K(\eta_2)^N,\,K>0,\,0\leq\eta_2<1,\,N=0,1,\dots,
\end{align}
where $\eta_2$ is the second-largest eigenvalue of $R$, and $(\eta_2)^N$ represents $\eta_2$ to the power of $N$.
\end{theorem}
\noindent{\bf Proof:} From ${^t}\bm{1}\bm\mu^N={^t}\bm{1}\bm\lambda^N-{^t}\bm{1}\widetilde{\bm\lambda}=0$, we have $\sum_{i=1}^n\mu_i^N=0$, and therefore $\mu_n^N=-\sum_{i=1}^{n-1}\mu_i^N$. Thus, $\mu_n^N$ is determined by $\mu_1^N,\dots,\mu_{n-1}^N$, so the effective variables are $\mu_1^N, \dots, \mu_{n-1}^N$. Let us define a vector $(\bm\mu^N)'$ by removing the $n$-th component $\mu_n^N$ from $\bm\mu^N$ as $(\bm\mu^N)'\triangleq{^t}(\mu_1^N,\dots,\mu_{n-1}^N)$, then the linear transformation $\bm\mu^{N+1}=R\bm\mu^N$ leads to the linear transformation
\begin{align}
(\bm\mu^{N+1})'=R'(\bm\mu^N)',\ \exists R'\in{\mathbb R}^{(n-1)\times(n-1)}.
\end{align}
Now, we find $R'$ and investigate its eigenvalues.

The transformation that removes the $n$-th component $\mu_n^N$ from $\bm\mu^N = {^t}(\mu_1^N,\dots,\mu_n^N)$ is given by
\begin{align}
\begin{pmatrix}\mu_1^N\\\vdots\\\mu_{n-1}^N\end{pmatrix}
=\begin{pmatrix}
1&0&\cdots&0&0\\
0&1&\cdots&0&0\\
\vdots&\vdots&\ddots&\vdots&\vdots\\
0&0&\cdots&1&0
\end{pmatrix}
\begin{pmatrix}\mu_1^N\\\vdots\\\mu_{n-1}^N\\\mu_n^N\end{pmatrix},
\end{align}
which can be written as
\begin{align}
(\bm\mu^N)'=E_0\bm\mu^N,\ E_0\triangleq\begin{pmatrix}
1&0&\cdots&0&0\\
0&1&\cdots&0&0\\
\vdots&\vdots&\ddots&\vdots&\vdots\\
0&0&\cdots&1&0
\end{pmatrix}
=\begin{pmatrix}I',\bm{0}'\end{pmatrix}\in{\mathbb R}^{(n-1)\times n},
\end{align}
where $I'\in\mathbb{R}^{(n-1)\times(n-1)}$ is the $(n-1)$-dimensional identity matrix, and $\bm{0}' \triangleq{^t}(0,\dots,0)\in\mathbb{R}^{n-1}$ is the zero vector of dimension $n-1$.

Next, define
\begin{align}
E_1\triangleq\begin{pmatrix}
1&0&\cdots&0\\
0&1&\cdots&0\\
\vdots&\vdots&\ddots&\vdots\\
0&0&\cdots&1\\
-1&-1&\dots&-1
\end{pmatrix}
=
\begin{pmatrix}I'\\-{^t}\bm{1}'\end{pmatrix}\in{\mathbb R}^{n\times (n-1)},
\end{align}
where $\bm{1}'={^t}(1,\dots,1)\in\mathbb{R}^{n-1}$. Then, we have
\begin{align}
&E_0E_1=\begin{pmatrix}I',\bm0'\end{pmatrix}\begin{pmatrix}I'\\-{^t}\bm{1}'\end{pmatrix}
=I'\in{\mathbb R}^{(n-1)\times(n-1)}\label{eqn:E0E1},\\
&E_1E_0=\begin{pmatrix}I'\\-{^t}\bm{1}'\end{pmatrix}\begin{pmatrix}I',\bm0'\end{pmatrix}
=I-\begin{pmatrix}{^t}\bm{0}\\\vdots\\{^t}\bm{0}\\{^t}\bm{1}\end{pmatrix}\in{\mathbb R}^{n\times n}.\label{eqn:E1E0}
\end{align}
Thus, from ${^t}\bm1\bm\mu^N=\bm0$, (\ref{eqn:Omega1=0}), and (\ref{eqn:E1E0}), we have
\begin{align}
E_1E_0\bm\mu^N=\bm\mu^N,\ E_1E_0\Omega=\Omega.\label{eqn:mu^NDelta_0Delta_1=mu^N}
\end{align}
Therefore, from (\ref{eqn:muN+1=muNR}) and (\ref{eqn:mu^NDelta_0Delta_1=mu^N}), we have $E_0\bm\mu^{N+1}=E_0R\bm\mu^N=E_0RE_1E_0\bm\mu^N$, which can be written as
\begin{align}
(\bm\mu^{N+1})'=R'(\bm\mu^N)',\ R'\triangleq E_0RE_1\in\mathbb R^{(n-1)\times(n-1)}.\label{eqn:mu^{N+1}'=(mu^N)'R'}
\end{align}
We now investigate the eigenvalues of $R'$. From (\ref{eqn:E0E1}) and (\ref{eqn:mu^{N+1}'=(mu^N)'R'}), we have
\begin{align}
R'=I'-E_0\Omega\,{^t}\Phi\Phi E_1\in{\mathbb R}^{(n-1)\times(n-1)}.\label{eqn:R'=Delta_1RDelta_0=}
\end{align}

Let
\begin{align}
W\triangleq E_0\Omega\,{^t}\Phi\Phi E_1\in{\mathbb R}^{(n-1)\times(n-1)},\label{eqn:Wdefinition}
\end{align}
and investigate the eigenvalues of $W$. From (\ref{eqn:mu^NDelta_0Delta_1=mu^N}), we have $E_1E_0\Omega\, {^t}\Phi\Phi=\Omega\,{^t}\Phi\Phi\in\mathbb{R}^{n \times n}$. From Lemma \ref{lem:eigenvaluesofXY=YX}, the eigenvalues of $\Omega\,{^t}\Phi\Phi\in{\mathbb R}^{n\times n}$ are obtained by adding one zero to the eigenvalues of $W=E_0\Omega\,{^t}\Phi\Phi E_1\in\mathbb{R}^{(n-1)\times(n-1)}$. In other words, the eigenvalues of $W$ are $\rho_2,\dots,\rho_n$, which are obtained by excluding $\rho_1=0$ from the $n$ eigenvalues $\rho_1=0,\rho_2,\dots,\rho_n$ of $\Omega\,{^t}\Phi\Phi$. Thus, from (\ref{eqn:0=beta1<beta2leqbetamleq1}), the eigenvalues of $W$ satisfy
\begin{align}
0<\rho_2\leq\dots\leq\rho_n\leq1.\label{eqn:0<rho2leqdotsleqrhonleq1}
\end{align}
Because the eigenvalues of $R'=I'-W$ are $\eta_i=1-\rho_i$ for $i=2,\dots,n$, we have from (\ref{eqn:0<rho2leqdotsleqrhonleq1})
\begin{align}
1>\eta_2\geq\dots\geq\eta_n\geq0.
\end{align}
Thus, the convergence rate of the recurrence formula (\ref{eqn:mu^{N+1}'=(mu^N)'R'}) is
\begin{align}
\|\left(\bm\mu^N\right)'\|\leq K_1\left(\eta_2\right)^N,\ K_1>0,\ N=0,1,\dots.\label{eqn:muN'<K_1theta2N}
\end{align}
By simple calculation, we know that $\|\bm\mu^N\|\leq\sqrt{n}\|\left(\bm\mu^N\right)'\|$, so from (\ref{eqn:muN'<K_1theta2N}), we obtain
\begin{align}
\|\bm\mu^N\|\leq K\left(\eta_2\right)^N,\ K>0,\ N=0,1,\dots,
\end{align}
which completes the proof of the theorem. \hfill$\square$

\subsection{Computational Complexity of Recurrence Formula (\ref{eqn:proposed1})}

Under the rank condition (\ref{eqn:rankdefinition}), $\bm\lambda^N$ defined by the recurrence formula (\ref{eqn:proposed1}) converges to $\widetilde{\bm\lambda}$ according to Theorem \ref{the:mainsecond}. We evaluate the computational complexity required for the error between $\bm\lambda^N$ and $\widetilde{\bm\lambda}$ to become less than $\epsilon$. The computational complexity is defined as the product of the number of operations per iteration $N$ in (\ref{eqn:proposed1}) and the number of iterations $N$ required to satisfy $\|\bm\lambda^N-\widetilde{\bm\lambda}\|<\epsilon$. The number of operations per iteration $N$ in (\ref{eqn:proposed1}) consists of $n^2$ multiplications and $n^2$ additions, resulting in a total of $2n^2$ operations. Next, for the value of $N$ such that $\|\bm\lambda^N-\widetilde{\bm\lambda}\|<\epsilon$, it suffices that $K(\eta_2)^N < \epsilon$ holds by Theorem \ref{the:mainsecond}, leading to the choice $N=\left(\ln(1/\epsilon)+\ln K\right)/\ln(1/\eta_2)$. Summarizing the above, the computational complexity of the recurrence formula (\ref{eqn:proposed1}) is given by $O\left(\kappa n^2 \ln(1/\epsilon)\right)$, where $\kappa \triangleq 1 / \ln(1/\eta_2)$. The reason for explicitly introducing $\kappa$ is that, depending on $\Phi$, $\eta_2$ can be arbitrarily close to 1, and $\kappa$ can therefore become arbitrarily large. An example of such $\Phi$ is provided in \ref{app:eta2example}.

We obtain the following theorem.
\begin{theorem}
\label{the:computationalcomplexity}
Assume the rank condition (\ref{eqn:rankdefinition}). Then the sequence $\bm\lambda^N$ defined by the recurrence formula (\ref{eqn:proposed1}) converges to the barycentric coordinate $\widetilde{\bm\lambda}$ of the equidistant point $\widetilde{Q}$ from $P^1,\dots,P^n$. The computational complexity required for the error between $\bm\lambda^N$ and $\widetilde{\bm\lambda}$ to become less than $\epsilon$ is given by
\begin{align}
O\left(\kappa n^2 \ln\left(\dfrac{1}{\epsilon}\right)\right),\label{eqn:computationalcomplexity}
\end{align}
where $\kappa \triangleq 1 / \ln(1/\eta_2)$ and $\eta_2$ is the second largest eigenvalue of $R$.
\end{theorem}
\begin{remark}
\rm If $\widetilde{\bm\lambda}\in\Delta^n$, i.e., $\widetilde{\lambda}_i\geq 0,\,i=1,\dots,n$, we have $\bm\lambda^\ast=\widetilde{\bm\lambda}$ (\cite[Lemma 2]{FGK03}). Therefore, in this case, $\bm\lambda^N$ converges to $\bm\lambda^\ast$, that is the barycentric coordinate of $Q^\ast$.
\end{remark}
\section{Numerical Examples (1)}
\label{sec:examples1}

This section provides numerical examples to investigate the convergence performance of the recurrence formula (\ref{eqn:proposed1}) under the rank condition (\ref{eqn:rankdefinition}).
%
%
\begin{example}
\label{exa:1}\rm 
Consider the case $d=2,n=3$, and let $\Phi=\begin{pmatrix} 1 & 3 & 2 \\ 0 & 0 & 2 \end{pmatrix}$, which satisfies the rank condition (\ref{eqn:rankdefinition}). The matrix $R$ of (\ref{eqn:proposed2}) and vector $\bm{c}$ of (\ref{eqn:proposed3}) are given by
\begin{align}
R&=\begin{pmatrix}
292/267 & 75/267 & 86/267\\
-17/267 & 216/267 & -30/267\\
-8/267 & -24/267 & 211/267
\end{pmatrix},\\
&\text{eigenvalues: }1,\,0.906,\,0.786,\\
\bm{c}&={^t}(-127/534,\ 65/534,\ 62/534).
\end{align}
Starting with the initial vector $\bm{\lambda}^0={^t}(1/3,1/3,1/3)$, we apply the recurrence formula (\ref{eqn:proposed1}) 100 times, obtaining the result $\bm{\lambda}^{100}={^t}(0.31249, 0.31250, 0.37499)$, which appears to converge to the barycentric coordinate $\bm{\lambda}^\ast={^t}(0.315,0.315,0.375)$ of the center $Q^\ast$ of the SEB.
\end{example}
%
%
\begin{example}
\label{exa:2}\rm 
For $d=2,n=3$, let $\Phi=\begin{pmatrix} 1 & 5 & 3 \\ 0 & 0 & 1 \end{pmatrix}$, which satisfies the rank condition (\ref{eqn:rankdefinition}). The matrix $R$ and vector $\bm{c}$ are
\begin{align}
R&=\begin{pmatrix}
315/285 & 30/57 & 59/171 \\
-14/285 & 43/57 & -25/171 \\
-16/285 & -16/57 & 137/171
\end{pmatrix},\\
&\text{eigenvalues: }1,\,0.980,\,0.680,\\
\bm{c}&={^t}(-31/114,\ 17/114,\ 14/114).
\end{align}
Starting with the initial vector $\bm{\lambda}^0={^t}(1/3,1/3,1/3)$, after 500 iterations, we obtain $\bm{\lambda}^{500}={^t}(1.24995,1.24994,-1.49990)$, which appears to converge to $\widetilde{\bm{\lambda}}={^t}(1.25,1.25,-1.5)$, that is the barycentric coordinate of the equidistant point $\widetilde{Q}$. However, in this example the barycentric coordinate of $Q^\ast$ is $\bm{\lambda}^\ast={^t}(1/2,1/2,0)$ and since $\bm{\lambda}^\ast\neq\widetilde{\bm{\lambda}}$, the sequence $\bm{\lambda}^N$ does not converge to $\bm{\lambda}^*$ as it is. We will show how to solve this problem in Example \ref{exa:4}.
\end{example}
%
%
\begin{example}
\label{exa:3}\rm 
Let $d=n$, and consider the points $P^i={^t}(0,\cdots,0,\stackrel{i\,\text{th}}{\stackrel{\vee}{1}},0,\cdots,0)\in\mathbb{R}^n,\ i=1,\cdots,n$, i.e., $\Phi=I\in \mathbb{R}^{n \times n}$, which satisfies the rank condition (\ref{eqn:rankdefinition}). In this case,
\begin{align}
R&=\dfrac{1}{n^2}\left\{(n^2-n)I+\bm{1}{^t}\bm{1}\right\},\\
\bm{c}&=\bm{0}.
\end{align}
We have $\bm{\lambda}^\ast={^t}(1/n,\dots,1/n)$. The characteristic polynomial of $\bm{1}{^t}\bm{1}\in\mathbb R^{n\times n}$ is $\det(\bm{1}{^t}\bm{1}-tI)=(-1)^n(t-n)t^{n-1}$ (\cite[p.56,\,1.2.P16]{hor}). From this, we see that the eigenvalues of $R$ are 1 with multiplicity 1, and $1-(1/n)$ with multiplicity $n-1$. Therefore, by Theorem \ref{the:mainsecond}, we have the convergence rate
\begin{align}
\limsup_{N\to\infty}\dfrac{1}{N}\log\|\bm\lambda^N-\bm\lambda^\ast\|\leq\log\left(1-\dfrac{1}{n}\right).\label{eqn:log(1-1m)}
\end{align}
For $n=29$, Figure \ref{fig:lognorm} shows the comparison between $(1/N)\log\|\bm{\lambda}^N-\bm{\lambda}^*\|$ and $\log(1-1/29)=-0.035$, demonstrating that the equality holds approximately.
\begin{figure}[H]
\centering
\begin{overpic}[width=0.7\columnwidth]{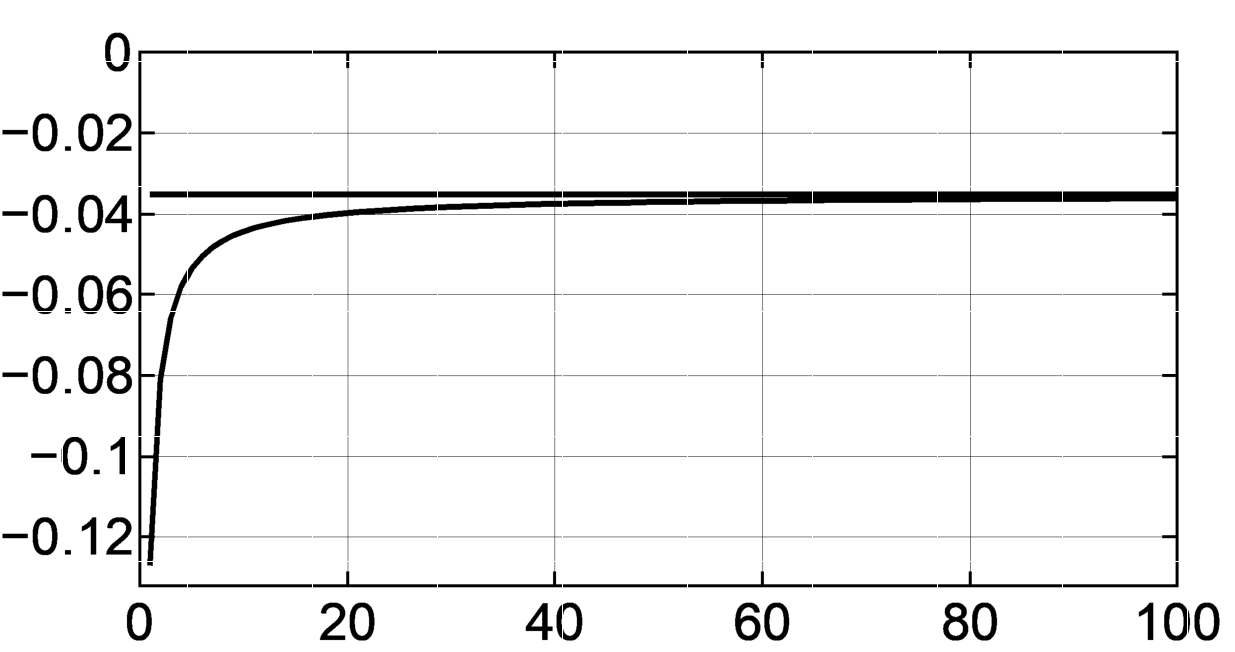}
\put(25,42.5){\vector(-1,-1){5}}
\put(26,43){$\log(1-1/29)=-0.035$}
\put(24,26){\vector(-1,2){3}}
\put(25,25){$(1/N)\log\|\bm\lambda^N-\bm\lambda^\ast\|$}
\put(53,-2){$N$}
\end{overpic}
\caption{The convergence rate of $\bm\lambda^N\to\bm\lambda^\ast$ in Example \ref{exa:3} for $n=29$.}
\label{fig:lognorm}
\end{figure}

In this example \ref{exa:3}, since $\bm\lambda^\ast=\widetilde{\bm\lambda}$, we can compare the proposed method with other conventional methods. We will compare below the run time of the proposed method with that of Welzl's method \cite{Welzl} and FGK's method \cite{FGK03}. The comparison is based on real-time measurements on a computer with AMD Ryzen 7 5800H CPU (clock speed 3.2 GHz base/4.4 GHz maximum).

The run time of the proposed method is the time (in seconds) until $\|\bm\lambda^N-\bm\lambda^\ast\|<10^{-6}$ is satisfied, starting with the initial vector $\bm\lambda^0={^t}(0.9,\,0.1/(n-1),\dots,0.1/(n-1))$. The run times of Welzl's and FGK's methods are the time (in seconds) until these methods determine that they have found the set of supporting points for the SEB and stop.

Figure \ref{fig:Welzl} shows a comparison of the run time between the proposed method and Welzl's method. The run time of Welzl's method increases rapidly with the dimension $n=d$. For example, at $n=20$, the time is about 2.2\,[s], but at $n=29$, it requires about 2200\,[s]. In contrast, the proposed method shows a gradual increase, but the run time remains stable around 0.9\,[s] for $10 \leq n \leq 29$, and it is faster than Welzl's method for $n \geq 20$.

Figure \ref{fig:FGK} shows a comparison of the run time between the proposed method and FGK's method. The run times of FGK's method are about 0.05\,[s], 2.9\,[s], 23\,[s] for $n=128,\,512,\,1024$, respectively. In comparison, the proposed method takes about 3\,[s], 150\,[s], 920\,[s] for $n=128,\,512,\,1024$, respectively. In this example, the proposed method is slower than FGK's method in this range of values for $n$.
\begin{figure}[H]
\centering
\begin{overpic}[width=0.7\columnwidth]{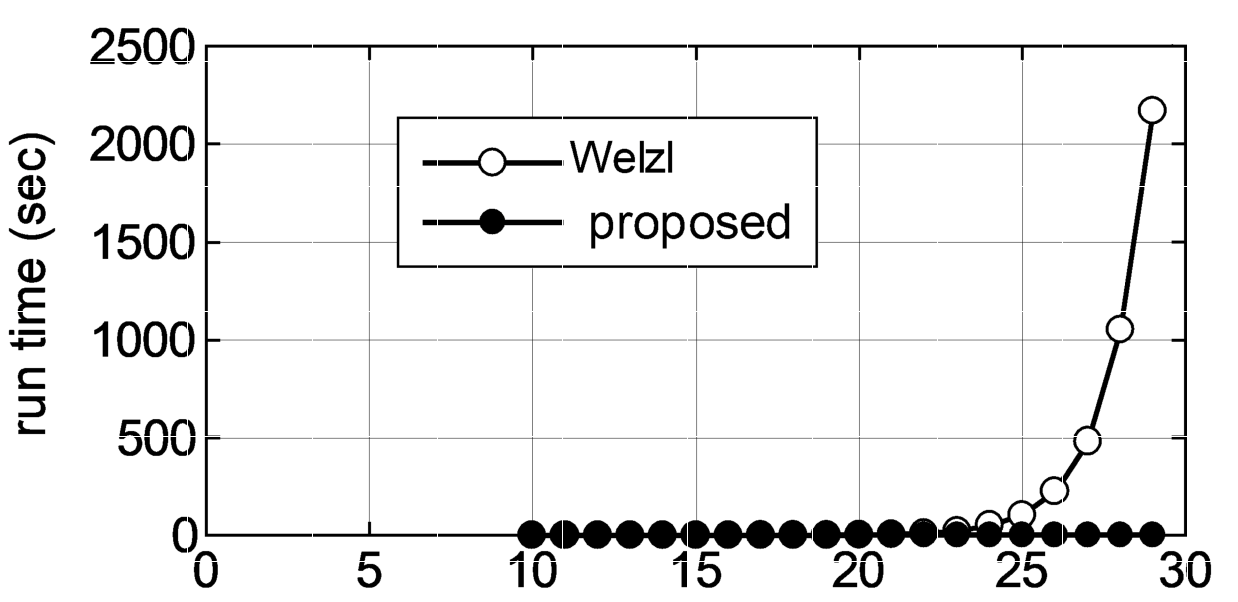}
\put(61,0){$n$}
\end{overpic}
\caption{Run time comparison with Welzl's Method \cite{Welzl}.}
\label{fig:Welzl}
\end{figure}
\begin{figure}[H]
\centering
\begin{overpic}[width=0.7\columnwidth]{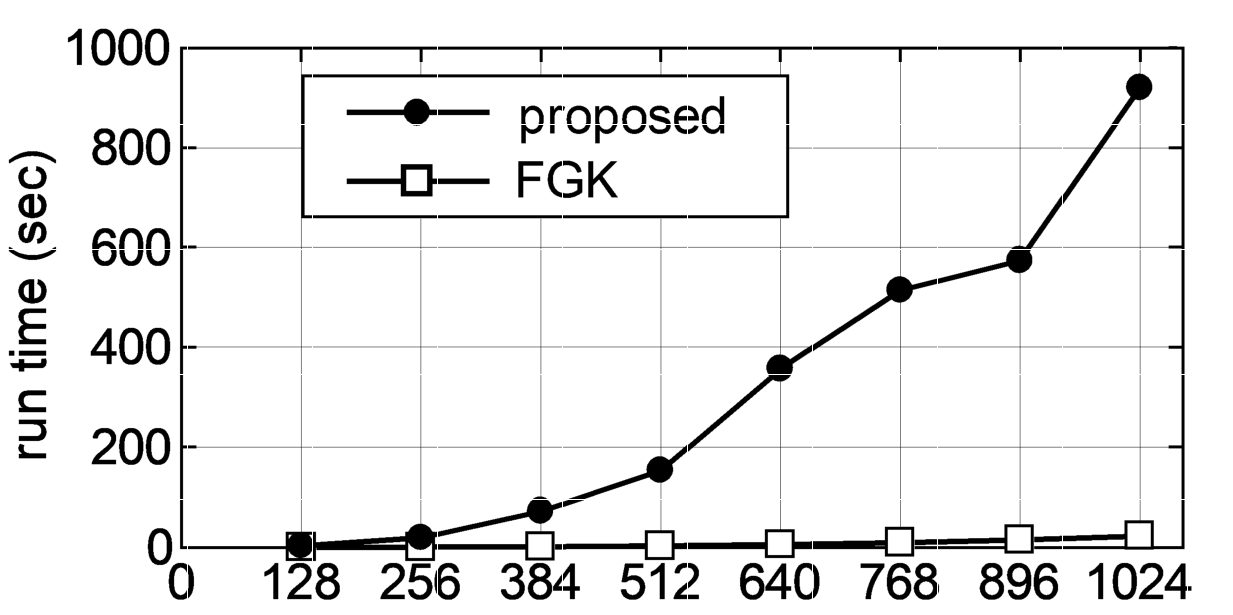}
\put(66,-2){$n$}
\end{overpic}
\caption{Run time comparison with FGK's Method \cite{FGK03}}
\label{fig:FGK}
\end{figure}
\end{example}
\section{Heuristic Algorithm}
\label{sec:heuristic}
The recurrence formula (\ref{eqn:proposed1}) is applicable to an arbitrary $\Phi$. In the case where $\Phi$ satisfies the rank condition (\ref{eqn:rankdefinition}), we have discussed in the preceding sections that the recurrence formula (\ref{eqn:proposed1}) converges to the barycentric coordinates $\widetilde{\bm\lambda}$ of the equidistant point $\widetilde{Q}$.

In this section, we consider cases where $\Phi$ satisfies the rank condition (\ref{eqn:rankdefinition}) but $\widetilde{\bm\lambda}\notin\Delta^n$, or where the rank condition (\ref{eqn:rankdefinition}) does not hold. For such cases, we propose a heuristic algorithm as a modification of the recurrence formula (\ref{eqn:proposed1}), and numerically evaluate its convergence performance.

Since the barycentric coordinate $\bm\lambda^\ast={^t}(\lambda^\ast_1,\dots,\lambda^\ast_n)$ of the center of the SEB must lie within $\Delta^n$, if $\lambda^N_i<0$ for some $N$ and $i$, it is considered that continuing the recurrence formula will not lead to convergence to $\bm\lambda^\ast$. Then, we propose the following heuristic algorithm.

\smallskip

\noindent\underline{Heuristic Algorithm}

\begin{figure}[htbp]
\centering
\begin{overpic}[width=0.5\columnwidth]{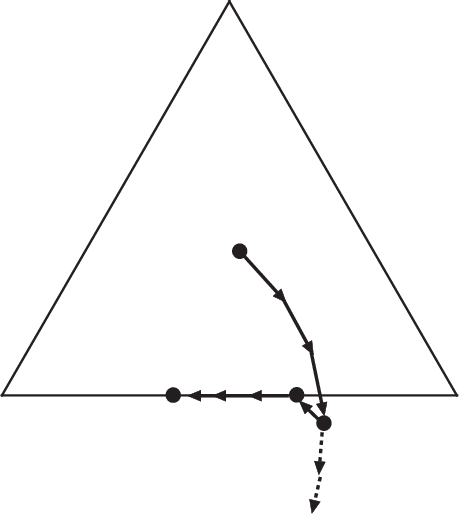}
\put(49,52){$\bm\lambda^0$}
\put(55,44){$\bm\lambda^1$}
\put(61,34){$\bm\lambda^{N-1}$}
\put(65,17){$\bm\lambda^N$}
\put(49.5,26){$\bm\lambda^{N+1}$}
\put(33,26.5){$\bm\lambda^\ast$}
\put(59,80){$\Delta^n$}
\end{overpic}
\caption{Heuristic Algorithm}
\label{fig:outofDelta}
\end{figure}
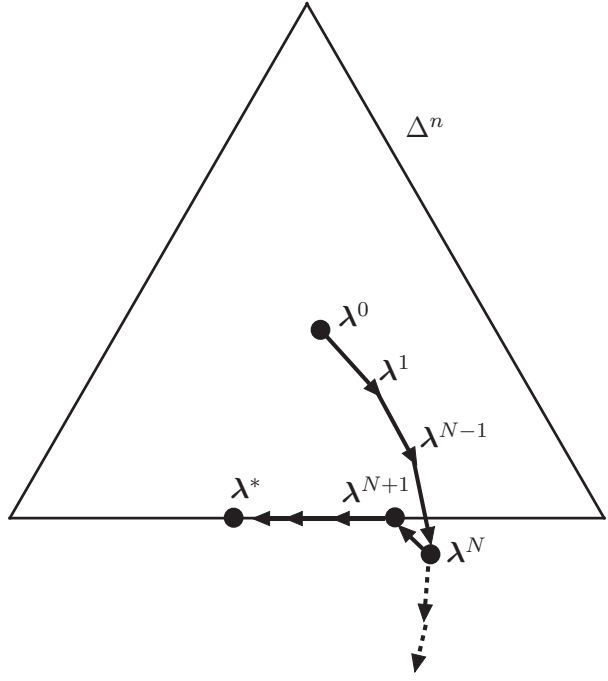

In the computation of $\bm\lambda^N={^t}(\lambda_1^N,\dots,\lambda_n^N)$ using the recurrence formula (\ref{eqn:proposed1}), if $\lambda_i^N<0$ for some $N$ and $i$, then exclude the point $P^i$ and reapply the recurrence formula (\ref{eqn:proposed1}) to the updated set of points $\{P^1,\dots,P^{i-1},P^{i+1},\dots,P^n\}$. See Figure \ref{fig:outofDelta}.
\section{Numerical Examples (2)}
\label{sec:numericalexamples2}
We have applied the heuristic algorithm for several $\Phi$'s, then there were cases where the correct $\bm{\lambda}^\ast$ was obtained and cases where it was not. Below, we present the case where the correct $\bm{\lambda}^\ast$ was obtained as Example \ref{exa:4} and the case where it was not as Example \ref{exa:5}.
%
%
\begin{example}
\label{exa:4}
\rm We apply the heuristic algorithm to $\Phi$ of Example \ref{exa:2}. When $N=9$, we have $\bm\lambda^9={^t}(0.578,0.440,-0.018)$, and since $\lambda_3^9=-0.018<0$, we exclude the corresponding point $P^3$. After excluding, we continue with the updated $R,\,\bm{c}$, and $\bm\lambda^9$ as
\begin{align}
R&=\begin{pmatrix}
14/13 & 5/13 \\
-1/13 & 8/13
\end{pmatrix},\\
&\text{eigenvalues: }1,\,0.692,\\
\bm{c}&={^t}(-3/13,\,3/13),\\
&\text{initial vector}\ \bm\lambda^9={^t}(0.568,\ 0.432).
\end{align}
Continuing with these, the true vector $\bm\lambda^\ast={^t}(0.5,0.5)$ is reached, in fact, at $N=42$, we have $\|\bm\lambda^{42}-\bm\lambda^\ast\|<10^{-6}$, resulting in the correct $\bm\lambda^\ast$.
\end{example}
%
%
\begin{example}
\label{exa:5}
\rm Let $d=2, n=4$, and
\begin{align}
\Phi=\left(P^1,P^2,P^3,P^4\right)
=\begin{pmatrix}0.441234 & -0.405275 & -0.499223 & 0.470587 \\
0.375473 & 0.405980 & 0.333663 & -0.422787
\end{pmatrix},
\end{align}
which does not satisfy the rank condition (\ref{eqn:rankdefinition}). We have $\bm\lambda^\ast={^t}(0.007718,0.000000,0.496774,0.495508)$, so $Q^\ast=\Phi\bm\lambda^\ast={^t}(-0.011416,-0.040841)$. Since $\lambda^\ast_1>0, \lambda^\ast_2=0,\lambda^\ast_3>0,\lambda^\ast_4>0$, the points $P^1, P^3, P^4$ support the SEB. However, the proposed heuristic algorithm excludes $P^1$ and $P^2$ during the iterative process and converges to the barycentric coordinate $\bm\lambda^\infty\triangleq{^t}(0,0,0.5,0.5)$, which leads to an incorrect center of the SEB at $Q^\infty\triangleq\Phi\bm\lambda^\infty={^t}(-0.014318,-0.044562)$. See Figure \ref{fig:circle}. But the difference between $Q^\infty$ and $Q^\ast$ is very small, so the heuristic algorithm can still be considered a practical approximation.
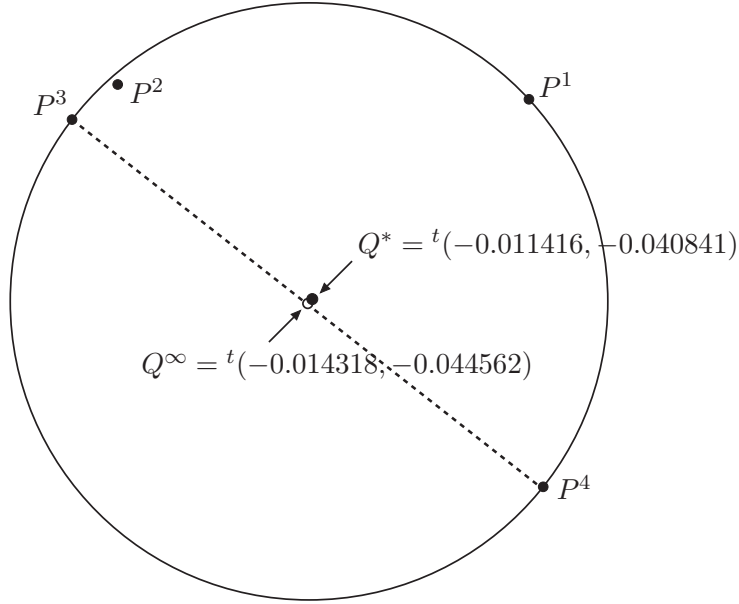
\begin{figure}[H]
\centering
\begin{overpic}[width=0.5\columnwidth]{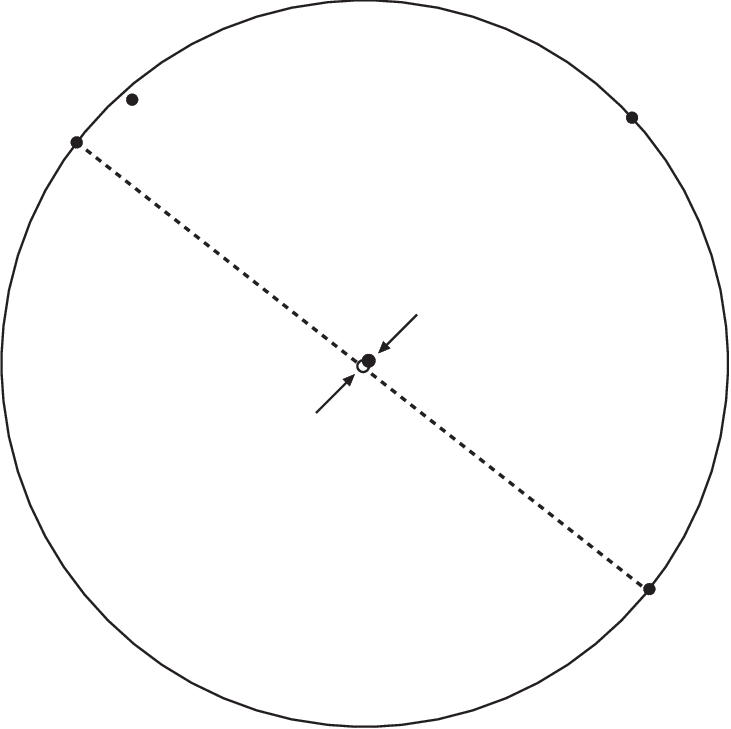}
\put(58,58){$Q^\ast={^t}(-0.011416, -0.040841)$}
\put(22,38){$Q^\infty={^t}(-0.014318, -0.044562)$}
\put(88,84){$P^1$}
\put(20,83){$P^2$}
\put(4,81){$P^3$}
\put(91,17){$P^4$}
\end{overpic}
\caption{An example where the correct $Q^\ast$ was not obtained.}
\label{fig:circle}
\end{figure}
\end{example}
%
%
\begin{example}
\label{exa:6}\rm Comparison of Run Time, Radius, and Center

We compare the run time, relative error in the radius of the SEB, and the error in the center of the SEB between the proposed heuristic algorithm and Welzl's method. The run time for Welzl's method is the one explained in Example \ref{exa:3}. Since Welzl's method gives the exact solution, the barycentric coordinate $\bm\lambda^\ast$ of the center, the radius $\sqrt{\Gamma}$, and the center $Q^\ast$ are obtained by Welzl's method.

In (i) below, the run times of the proposed method are shown, which are measured as the time (in seconds) until $\|\bm\lambda^N-\bm\lambda^\ast\|<10^{-6}$. In (ii), the values of $|\sqrt{\Gamma^N}-\sqrt{\Gamma}|/\sqrt{\Gamma}$ are shown, where $\Gamma^N\triangleq\max_{1\leq i\leq n}\|P^i-Q^N\|^2$. In (iii), the values of $\|Q^N-Q^\ast\|$ are shown.

Specifically, the evaluation is conducted using the following numerical examples.

$n$ points are selected uniformly at random from the $d$-dimensional unit hypercube. For each combination of $d$ and $n$ listed below, $16$ different point sets are generated. All of these point sets do not satisfy the rank condition (\ref{eqn:rankdefinition}).

\medskip

case 1\,:\,$d=16, \ n=128$
                
case 2\,:\,$d=32, \ n=128$
                
case 3\,:\,$d=16, \ n=256$
                
case 4\,:\,$d=32, \ n=256$

\medskip

(i) Run Time (in seconds) (average per case)
\begin{table}[H]
\centering
\begin{tabular}{|r|r|r|}
\hline
& Welzl & proposed\\
\hline
case 1 & 0.001 & 1.663\\
\hline
case 2 & 0.684 & 2.427\\
\hline
case 3 & 0.001 & 3.064\\
\hline
case 4 & 0.441 & 4.507\\
\hline
\end{tabular}
\end{table}

The run times of Welzl's method are smaller than those of the proposed method, but it's not as though the proposed method is particularly time-consuming. The run time of Welzl's method grows larger in higher-dimensional cases, i.e., cases 2 and 4.

\bigskip

(ii) Relative Error in the Radius of the SEB (average per case)

\begin{table}[H]
\centering
\begin{tabular}{|r|r|}
\hline
case 1 & 8.97E-04\\
\hline
case 2 & 2.34E-03\\
\hline
case 3 & 3.18E-03\\
\hline
case 4 & 3.15E-03\\
\hline
\end{tabular}
\end{table}

(iii) Error in the Center of the SEB (average per case)

\begin{table}[H]
\centering
\begin{tabular}{|r|r|}
\hline
case 1 & 8.97E-04\\
\hline
case 2 & 2.34E-03\\
\hline
case 3 & 3.18E-03\\
\hline
case 4 & 3.15E-03\\
\hline
\end{tabular}
\end{table}

The errors in (ii) and (iii) are small, so the proposed method is sufficiently practical for most cases.
\end{example}
\section{Conclusion and Future Work}

In this paper, we developed an iterative algorithm for computing the center of the SEB (smallest enclosing ball) of $n$ points $P^1,\dots,P^n$ in $d$-dimensional Euclidean space $\mathbb{R}^d$. The resulting theorem shows that when $\Phi=(P^1,\dots,P^n)$ satisfies the rank condition (\ref{eqn:rankdefinition}), the proposed recurrence formula (\ref{eqn:proposed1}) converges exponentially to the barycentric coordinate $\widetilde{\bm\lambda}$ of the equidistant point $\widetilde{Q}$ from $P^1,\dots,P^n$. If $\widetilde{\bm\lambda}\in\Delta^n$, then $\widetilde{\bm\lambda}=\bm\lambda^\ast$ holds, where $\bm\lambda^\ast$ is the barycentric coordinate of the center $Q^\ast$ of the SEB. Thus, in this case, $\bm\lambda^N$ converges to $\bm\lambda^\ast$. Moreover, the convergence rate was evaluated based on the eigenvalues of the matrix $R$ of (\ref{eqn:proposed1}). We also applied our heuristic algorithm when the rank condition (\ref{eqn:rankdefinition}) is not satisfied, and conducted numerical experiments to study the convergence performance. The proposed method has some advantages and some disadvantages compared to well-established methods such as Welzl's\cite{Welzl} and FGK's\cite{FGK03} algorithms. But overall, since the proposed method can be computed using a very simple formula, it is considered sufficiently practical.

In the future, we aim to further improve the algorithm and develop a version that guarantees convergence to $\bm\lambda^\ast$ for any point set $\Phi$.

\appendix
\def\thesection{Appendix \Alph{section}}
\section{Proof of Theorem \ref{the:Gamma=maxJlambda}}  
\label{app:A}

To prove Theorem \ref{the:Gamma=maxJlambda}, we prepare several lemmas.
\begin{lemma}
\label{lem:compactset}
We have
\begin{align}
\min_{Q\in{\mathbb R^d}}\max_{1\leq i\leq n}\|P^i-Q\|^2=\min_{Q\in\Phi\Delta^n}\max_{1\leq i\leq n}\|P^i-Q\|^2.\label{eqn:minislimited}
\end{align}
\end{lemma}

\noindent{\bf Proof:} (Note that $\Phi \Delta^n$ is defined in (\ref{eqn:DeltamPhidefinition}).) We  prove (\ref{eqn:minislimited}) by contradiction. Let $Q=Q^\ast$ achieve the minimization of the left-hand side of (\ref{eqn:minislimited}). Since $\Phi\Delta^n$ is a bounded convex closed set, if $Q^\ast\notin\Phi\Delta^n$ there exists a hyperplane $\pi$ that separates $\{P^1,\dots,P^n\}$ and $Q^\ast$, i.e., $\{P^1,\dots,P^n\}$ and $Q^\ast$ lie on opposite sides of $\pi$. Let $Q_0$ be the point on $\pi$ closest to $Q^\ast$. Then, $\|P^i-Q_0\|<\|P^i-Q^\ast\|$ for $i= 1,\dots,n$, which contradicts the fact that $Q^\ast$ is the center of the SEB. \hfill$\square$
\begin{lemma}
\label{lem:continuousmax}
For non-negative real numbers $x_i\geq 0,\,i=1,\dots,n$, we have
\begin{align}
\ds\max_{1\leq i\leq n}x_i=\ds\max_{\bm\lambda\in\Delta^n}\ds\sum_{i=1}^n\lambda_ix_i.\label{eqn:maxmax}
\end{align}
\end{lemma}
\noindent{{\bf Proof:} Without loss of generality, assume $x_1\geq x_2\geq\dots\geq x_n\geq0$. Then, specifically for $\bm{\lambda}^\dagger=(\lambda_1^\dagger,\dots,\lambda_n^\dagger)\triangleq(1,0,\dots,0)$, we have that the right-hand side $\geq\sum_{i=1}^n\lambda_i^\dagger x_i=x_1=$ the left-hand side. Also, we have that the right-hand side $\leq\max_{\bm{\lambda}\in\Delta^n}\sum_{i=1}^n\lambda_ix_1=x_1=$ the left-hand side. \hfill $\square$

\begin{lemma}
\label{lem:minmaxexchange}
{\rm (See \cite[Theorem 1.5.1]{kar})} Let $A,B$ be bounded closed convex sets in $\mathbb R^m$ and in $\mathbb R^n$, respectively. Let $f(\bm{x},\bm{y})$ be a real valued function of $\bm{x}\in A$ and $\bm{y}\in B$. If $f(\bm{x},\bm{y})$ is continuous, convex in $\bm{y}\in B$ for each $\bm{x}\in A$, and concave in $\bm{x}\in A$ for each $\bm{y}\in B$, then,
\begin{align}  
\min_{\bm{y}\in B}\max_{\bm{x}\in A}f(\bm{x},\bm{y})=\max_{\bm{x}\in A}\min_{\bm{y}\in B}f(\bm{x},\bm{y}).
\end{align}
\end{lemma}

\bigskip

\noindent{\bf Proof of Theorem \ref{the:Gamma=maxJlambda}:} (Note that $\Gamma$ is defined in (\ref{eqn:Gammadefinition2}).) We have
\begin{align}
\Gamma&=\min_{Q\in\Phi\Delta^n}\max_{1\leq i\leq n}\|P^i-Q\|^2\ \ \text{(from Lemma \ref{lem:compactset})}\\
&=\min_{Q\in\Phi\Delta^n}\max_{\bm\lambda\in\Delta^n}\sum_{i=1}^n\lambda_i\|P^i-Q\|^2\ \ \text{(from Lemma \ref{lem:continuousmax})}\\
&=\max_{\bm\lambda\in\Delta^n}\min_{Q\in\Phi\Delta^n}\sum_{i=1}^n\lambda_i\|P^i-Q\|^2\ \ \text{(from Lemma \ref{lem:minmaxexchange})}\\
&=\max_{\bm\lambda\in\Delta^n}\min_{Q\in\Phi\Delta^n}\{J(\bm\lambda)+\|\Phi\bm\lambda-Q\|^2\}\ \ \text{(from Lemma \ref{lem:sumlambdaiPi-Q2})}\\
&=\max_{\bm\lambda\in\Delta^n}J(\bm\lambda)\ \ (\text{by substituting }Q=\Phi\bm\lambda).
\end{align}
\hfill$\square$

\section{Example where the Second Largest Eigenvalue $\eta_2$ of $R$ Becomes Arbitrarily Close to 1}  
\label{app:eta2example}

Let $d=2$, $n=3$, and $\Phi=\begin{pmatrix}-1&p&p\\0&q&-q\end{pmatrix}$, that is, $P^1={^t}(-1,0),\,P^2={^t}(p,q),\,P^3={^t}(p,-q)$, where $p^2+q^2=1$, $0<p<1$, and $q>0$. See Figure \ref{fig:PositionsofP1P2P3}.

\begin{figure}[H]
\centering
\begin{overpic}[width=0.45\columnwidth]{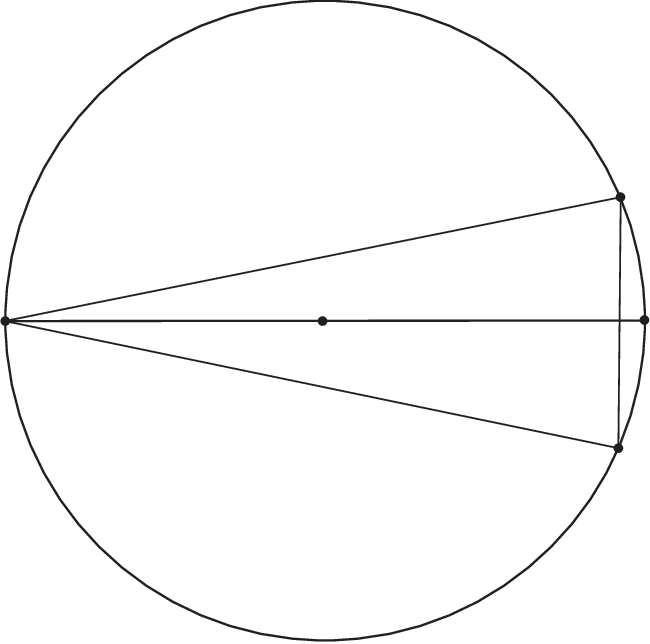}
\put(-34,48){$P^1={^t}(-1,0)$}
\put(45,44){$O$}
\put(98,68){$P^2={^t}(p,q)$}
\put(98,28){$P^3={^t}(p,-q)$}
\put(101,48){${^t}(1,0)$}
\end{overpic}
\caption{Positions of $P^1,P^2,P^3$.}
\label{fig:PositionsofP1P2P3}
\end{figure}
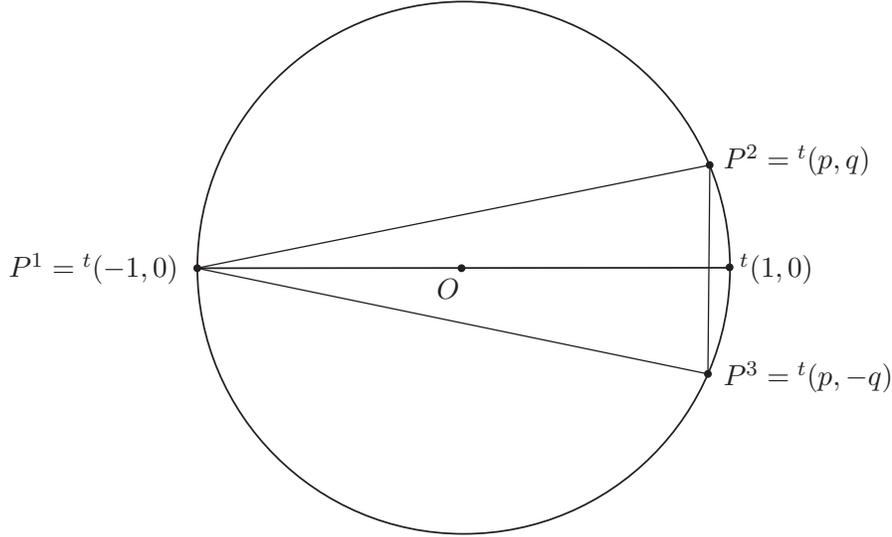

The matrix $\Phi$ satisfies the rank condition (\ref{eqn:rankdefinition}). Furthermore, since the triangle $\triangle P^1P^2P^3$ is acute, we have $\widetilde{\bm\lambda}\in\Delta^3$, and thus $\bm\lambda^\ast=\widetilde{\bm\lambda}$. Therefore, it follows from Theorem \ref{the:mainsecond} that $\bm\lambda^N\to\bm\lambda^\ast$ as $N\to\infty$.

Now, consider the eigenvalues $\rho_2,\rho_3$ of the matrix $W=E_0\Omega\,{^t}\Phi\Phi E_1\in\mathbb{R}^{2\times2}$ corresponding to the above $\Phi$. From (\ref{eqn:0<rho2leqdotsleqrhonleq1}), we have $0<\rho_2\leq\rho_3\leq 1$, hence $0<(\rho_2\rho_3)^2\leq(\rho_2)^2\leq\rho_2\rho_3$ holds. Since $\det W=\rho_2\rho_3$, we have
\begin{align}
0<\left(\det W\right)^2\leq(\rho_2)^2\leq\det W.
\end{align}
Therefore, $\rho_2 \doteqdot 0$ is equivalent to $\det W \doteqdot 0$. Note that $\eta_2=1-\rho_2$, and a direct calculation yields
\begin{align}
\det W=\frac{4}{27}(1-p)(1+p)^3,\ 0<p<1,
\end{align}
so, it follows that $\eta_2\doteqdot1$ is equivalent to $p\doteqdot1$. Hence, by placing $P^2$ and $P^3$ arbitrarily close to the point ${^t}(1, 0)$, we obtain $p\doteqdot1$, and thus $\eta_2$ becomes arbitrarily close to 1. \hfill$\square$

\end{document}